\documentclass[10pt,aps,pra,twocolumn,superscriptaddress,floatfix,nofootinbib]{revtex4-1}
\makeatletter
\def\@bibdataout@aps{%
 \immediate\write\@bibdataout{%
  @CONTROL{%
   apsrev41Control,author="08",editor="1",pages="0",title="0",year="1",eprint="1"%
  }%
 }%
 \if@filesw
  \immediate\write\@auxout{\string\citation{apsrev41Control}}%
 \fi
}%
\makeatother 
\usepackage{lipsum}
\usepackage{amsmath}
\usepackage{amsfonts, amssymb, bbm, braket, accents}
\usepackage{graphicx}   
\usepackage[usenames,dvipsnames]{color}
\usepackage[makeroom]{cancel}
\usepackage{soul} 

\newcommand\etal{{\em et al.}}

\newcommand{\ain}{a_{{\rm in}}}
\newcommand{\aout}{a_{{\rm out}}}
\newcommand{\bin}{b_{{\rm in}}}
\newcommand{\bout}{b_{{\rm out}}}

\newcommand{\opin}[1]{#1_\textrm{in}}
\newcommand{\opout}[1]{#1_\textrm{out}}
\newcommand{\oppl}[2]{#1_{+}^{(#2)}}
\newcommand{\opmi}[2]{#1_{-}^{(#2)}}
\newcommand{\opz}[2]{#1_{z}^{(#2)}}
\newcommand{\conj}[1]{{#1}^{*}}

\newcommand{\dg}{^\dagger}
\newcommand{\smallfrac}[2]{\mbox{$\frac{#1}{#2}$}}
\newcommand{\half}{\smallfrac{1}{2}}
\newcommand{\ketbra}[2]{\left\vert{#1}\right\rangle\!\left\langle{#2}\right\vert}

\newcommand{\Id}{\openone}

\newcommand{\ip}[2]{\left\langle{#1}\right|\left.{#2}\right\rangle}

\newcommand{\op}[2]{\left |{#1}\right\rangle\! \!\left \langle {#2}\right |}

\usepackage[breaklinks=true]{hyperref}
\hypersetup{
  colorlinks   = true, 
  urlcolor     = blue, 
  linkcolor    = blue, 
  citecolor   = red 
}
\usepackage{amsthm}
\usepackage{cleveref} 
\crefformat{equation}{Eq.~(#2#1#3)} 
\Crefformat{equation}{Equation~(#2#1#3)}
\crefformat{figure}{Fig.~#2#1#3}
\crefrangeformat{equation}{Eqs.~#3(#1)#4--#5(#2)#6}

\begin{document}

\title{Two photons co- and counter-propagating through $N$ cross-Kerr sites}

\author{Daniel J. Brod}
\email{dbrod@perimeterinstitute.ca}
\affiliation{Perimeter Institute for Theoretical Physics, 31 Caroline St. N, Waterloo, Ontario, Canada N2L 2Y5}

\author{Joshua Combes}
\email{jcombes@perimeterinstitute.ca}
\affiliation{Institute for Quantum Computing and Department of Applied Mathematics, University of Waterloo, Waterloo, ON, Canada}
\affiliation{Perimeter Institute for Theoretical Physics, 31 Caroline St. N, Waterloo, Ontario, Canada N2L 2Y5}

\author{Julio Gea-Banacloche}
\email{jgeabana@uark.edu}
\affiliation{Department of Physics, University of Arkansas, Fayetteville, AR 72701, USA}

\date{\today}


\begin{abstract}
A cross-Kerr interaction produces a phase shift on two modes of light proportional to the number of 
photons in both modes, and is sometimes called cross-phase modulation. Cross-Kerr nonlinearities have 
many applications in classical and quantum 
nonlinear optics, including the possibility of a deterministic and all-optical controlled-phase gate. 
We calculate the one- and two-photon S-matrix for fields propagating in a medium where the cross-Kerr 
interaction is spatially distributed at discrete interaction sites comprised of atoms.
For the interactions considered, we analyze the cases where the photons co-propagate and counter-propagate 
through the medium and give a physical interpretation to the differences between the two cases.
Finally, we obtain the S-matrix in the limit of infinitely long chains, showing that it corresponds to 
a perfect controlled-phase operation.
\end{abstract}


\maketitle

\section{Introduction}\label{sec:intro}
Cross-Kerr nonlinearities have been suggested as solutions to a variety of optical quantum information processing tasks since Milburn's 1989 proposal of a quantum optical Fredkin gate \cite{Milb89}. In the field of quantum computing, strong Kerr nonlinearities ($\chi^{(3)}$) have been suggested for circuit-model \cite{ChuaYama95} and measurement based quantum computing \cite{HutcMilb04}, as have weak Kerr nonlinearities \cite{MunrNemoSpil05,LouiNemoMunr07}. Beyond quantum computing, cross-Kerr nonlinearities have also been proposed for all optical switching \cite{SandMilb92}, generation of entangled coherent states \cite{SandRice99}, quantum teleportation \cite{VitaFortTomb00}, Fock state conversion \cite{ClauKnolWels02}, entanglement distillation \cite{DuanGiedCira00,ClauKnolWels03, FiurMistFili03,MenzKoro06}, nonlinear quantum metrology \cite{BoixDattDavi08}, and quantum nondemolition (QND) detection of photon number \cite{MunrNemoBeau05}. Broadly speaking, one of the advantages of cross-Kerr interactions is that they are  photon-number-preserving, making them suitable for several of these applications. 

The goal of this paper is to present a full theoretical description of the interaction between two single-photon wave packets mediated by a network of cross-Kerr interaction sites. We use input-output theory and the theory of cascaded open quantum systems to model the interaction sites, which can be realized physically in a number of equivalent ways, such as cavities containing $\chi^{(3)}$ Kerr media, three- or four-level atoms, pairs of atoms with dipole coupling, etc. We then use this formulation to solve the one- and two-photon transport problem, where the photon wave packets can be either co- or counter-propagating.
Although this work is largely motivated by the design of a CPHASE gate (also often called a controlled-phase or controlled-$Z$ gate) that takes into account the fully-multimode nature of photon wave-packets, it also has more general applications to the quantum theory of cross-Kerr nonlinearities and quantum nonlinear optics. 

Our solutions to the few-photon transport problem are given in terms of S-matrices. The S-matrix is a unitary matrix that connects asymptotic input and output field states, and is a canonical way to characterize the action of a medium on light. The few-photon S-matrix has been studied for a long time in quantum optics \cite{DeutChiaGarr92,DeutChiaGarr93}, and there has been renewed interest since the work of Shen and Fan~\cite{ShenFan05}.  Indeed, the scattering problem for small photon number has applications and is interesting in and of itself. There is a vast literature on the subject, some of which is summarized in Ref.~\cite{RoyWilsFirs16}. 

We now briefly summarize the literature that is particularly related to our work, namely works concerning photon scattering in Kerr media and spin chains. The early investigations by Deutsch \etal~\cite{DeutChiaGarr92,DeutChiaGarr93} considered an atomic vapor that was modelled as continuous spatially-distributed self-Kerr medium, and solved the 
two-photon transport problem. 
Recently, a few works  considered coupling two fields at a single spatial site via a cavity-mediated self-Kerr interaction 
\cite{LiaoLaw10,XuRephFan13}, as well as one-dimensional arrays of coupled cavities with a self-Kerr medium driven by coherent \cite{BielMazzCaru15} or Fock~\cite{LeeNohSche15}  states.

Regarding scattering off cascaded spin chains, Roy~\cite{Roy13} considered an input-output channel coupled to the endpoints of a chain of three two-level systems coupled by hopping terms. Fang \etal~ \cite{FangZhenBara14} studied scattering of photons by three two-level systems  coupled to a waveguide supportting left and right propagating modes, and later~\cite{FangBara15} generalized this to $N$ two-level systems. Chiral and non-chiral single-photon transport through one-dimensional waveguide with a linear atomic chain side-coupled to the wave guide were considered in Refs.~\cite{LiaoZengZhu15,Koca16,LiaoNhaZuba16}.

\subsection{Further motivation and structure of the paper}\label{sec:struc}

There are two notable problems with most of the proposed uses of cross-Kerr nonlinearities of Refs.\ \cite{Milb89,ChuaYama95,HutcMilb04,MunrNemoSpil05,LouiNemoMunr07,SandMilb92,SandRice99,VitaFortTomb00,ClauKnolWels03,DuanGiedCira00,ClauKnolWels02,FiurMistFili03,MenzKoro06,BoixDattDavi08,MunrNemoBeau05}. The first is that they treat the field as a single frequency mode, whereas for propagating fields a full multimode analysis is more suitable, since frequency mixing within and between modes can ruin operation of devices. Based on such an analysis, Shapiro~\cite{Shap06} and Gea-Banacloche~\cite{GeaBana10} have argued that single-photon cross-Kerr nonlinearities
cannot straightforwardly be used to construct a CPHASE gate with any useful fidelity.  Similar conclusions were drawn for QND detection of photon number at microwave frequencies \cite{FanKockComb13}. The second problem is that, historically, experimentally-demonstrated cross-Kerr nonlinearities have been very weak~\cite{VenkSahaGaet13}, and experimentalists need to go to extraordinary lengths to observe an effect in the quantum domain \cite{FeizHallDmoc15}.

In some instances, however, the issues that arise from the multimode treatment can be overcome with careful analysis and some elaborate tricks. For example, a proposal for a CPHASE gate by Chudzicki et al.~\cite{ChudChuaShap13} considers a series of  localized $\chi^{(3)}$ interaction sites interleaved with active error correction.  Similar results were  obtained for QND photon-number detection in Refs.~ \cite{SathTornKock14,FanJohaComb14}.  
Additionally, in recent years large cross-Kerr nonlinearities have been demonstrated, for example, in cavity QED \cite{TurcHoodLang95a} ($\approx\!0.28$ radians per photon), atomic ensembles \cite{BeckHossDuan15} ($\approx\!1.05$ radians per photon), and single artificial atoms \cite{HoiKockPalo13} ($\approx\!0.35$ radians per photon). Both of these positive outcomes suggest  it may be valuable to consider alternative realizations, at the microscopic level, of cross-Kerr interaction media for single photons.

As one particular application of our results, we use the scattering matrix that describes two photons counter-propagating in an $N$-site chain to show that, in the limit where $N$ is large and the photons are spectrally narrow, a CPHASE gate can be performed with near-unit fidelity. In a related paper \cite{BrodComb16b}, two of us analyze the performance of our proposed setup when $N$ is small, showing that very high fidelities can be obtained for just a few dozen interactions sites, giving hope that our proposal may lead to an experimentally-feasible construction.

The calculations in this paper are detailed, so in \Cref{sec:notation} we set some notation and conventions. In \Cref{sec:inputoutput}, we give a brief overview of input-output theory and the cascaded theory of open quantum system, in particular in the form of the SLH formalism. This formalism is useful to obtain our differential equation used in the remained of the paper, but the familiar reader can skip this Section. In \Cref{sec:single} we determine the single- and two-photon S-matrices when there is a Kerr interaction at a single site in space, making heavy use of the methodology developed by Fan, Kocaba\c{s}, and Shen~\cite{FanKocaShen10}. Although these results have previously been obtained using a different formalism in \cite{ViswGeaB15}, we use this Section as an opportunity to introduce our mathematical ingredients in a simple setting. Our general philosophy is to outline the crucial steps of the scattering calculations and put the details in an appendix. In \Cref{sec:single} we also discuss several different physical implementations for which our results apply.

In \Cref{sec:2coprop} and \Cref{sec:2counter} we determine the one- and two-photon S-matrices when the photons are co- and counter-propagating, respectively, and there are two interaction sites. This investigation is inspired by related work on QND photon detection at microwave frequencies, where it was shown that adding more systems can improve detection efficiency \cite{SathTornKock14}. We generalize the counter-propagating analysis to $N$ interaction sites in \Cref{sec:Ncounter}, and then investigate the $N \rightarrow \infty$ limit in \Cref{sec:continuumcounter}. We conclude with a discussion of the similarities and differences of co- and counter-propagating configurations and its possible consequences for various applications.

\subsection{Notation and conventions}\label{sec:notation}

In this paper, we will consider two input-output fields which couple independently to separate physical systems. Thus, we denote the input-output fields by $\ain$ or $\bin$ and $\aout$ or $\bout$, respectively. The input field $\ain$ couples to cavity or atomic operators labelled by $a$ or $A$, while the input field $\bin$ couples to operators labelled by $b$ or $B$. Since none of our Hamiltonians transfer excitations between modes $a$ and $b$ in any way, statements such as ``there are no excitations in mode $a$'' should be understood to mean that there are no excitations in the collection of (i) external mode $\ain$, plus (ii) any cavity modes $a$ or atoms $A$. Finally, when fields are coupled to a chain of multiple similar systems, we call each ``unit cell''  an interaction site.

We denote frequency eigenstates in the infinite past by $\ket{\nu^{+}}$, and those in the infinite future by $\ket{\omega^{-}}$. When there are two modes of the field, $a$ and $b$, they shall be identified by their respective indices, e.g.\ $\ket{\nu_a^+ \nu_b^+}$. To simplify the notation, we usually omit vacuum states of unoccupied modes, e.g.\ $\ket{\nu_a^+}:=\ket{\nu_a^+ 0_b}$, unless there is risk of ambiguity. It should be pointed out that, given our physical setting, all states of the type $\ket{\nu_a^+}$ or $\ket{\omega_a^{-} \omega_b^{-}}$ also include implicitly an atomic ground state for all the atoms in the system. We denote the Pauli-$Z$ matrix on atom $A$ by $A_z := \ketbra{0}{0} - \ketbra{1}{1}$ and $B_z$ for atom $B$. The atomic states are represented by $\ket{0}$ and $\ket{1}$, but should just be understood as the standard ground and excited states, $\ket{g}$ and $\ket{e}$, respectively. We define the atomic ladder operators on atom $A$ as $A_- = \ketbra{0}{1}$, $A_+=A_-\dg = \ketbra{1}{0}$ and similarly for atom $B$. 

When there are multiple interaction sites the atomic operators become $\opz{A}{i}$ and $\opmi{A}{i}$, for $A_z$ and $A_-$ acting on atom $A$ at site $i$. Finally we point out that we made a few nonstandard choices, such as writing the atomic interaction as $(\Id-A_z )\otimes(\Id-B_z )$ rather than just $A_z  \otimes B_z $, to simplify comparison between cavity- and atom-mediated interactions. It is clear that these conventions can be mapped to more standard ones by unimportant global or local energy redefinitions. Finally, we will omit the tensor product sign throughout the paper whenever there is no risk of ambiguity.

\section{Theoretical Background}\label{sec:inputoutput}

In this section, we describe some of the theoretical background that is useful in obtaining our main results. In \cref{sec:IOT} we outline the basics of input-output theory, which describes the interaction between a quantum system and an external field in terms of the incoming and outgoing field operators. However, in our paper we actually want to solve the scattering problem for several different networks of connected quantum systems. Although it is possible, in principle, to do this by identifying the outputs of some quantum systems as inputs of others, this quickly becomes impractical. To facilitate this description, in \cref{sec:SLH} we describe the SLH formalism for composing and cascading of open quantum systems. This formalism consists of a set of algebraic relations that allows us to easily describe a network of quantum systems as a single larger quantum system, from which the relevant differential equation can be obtained directly, without the need to perform manipulations by hand. This Section can be safely skipped by the reader familiar with this subject.

\subsection{Input-Output theory}\label{sec:IOT}

The starting point for our analysis is the Gardiner-Collett Hamiltonian \cite{CollGard84,GardColl85}. The 
simplest form of their Hamiltonian models a single arbitrary system (e.g. an atom) interacting with a one-dimensional
chiral field. The Gardiner-Collett Hamiltonian is
\begin{align}
H_{\rm T}&= H_{\rm sys}+H_{\rm B}+H_{\rm Int},\nonumber \\
H_{\rm B} &=  \hbar \int d\omega\, \omega b\dg(\omega)b(\omega), \label{eq:GCH}\\
H_{\rm Int} &= i\hbar \sqrt{\gamma}\int d\omega  [L b\dg(\omega)-L\dg b(\omega)], \nonumber
\end{align}
where the field operators obey $[b(\omega),b\dg(\omega')] = \delta(\omega-\omega')$, $L$ is an operator on the 
system, and henceforth $\hbar=1$. Using this Hamiltonian and the machinery developed in \cite{YurkDenk84,CollGard84,GardColl85}
one can find equations of motion for an arbitrary system operator $X$,
\begin{align} \label{eq:fieldop1}
 \partial_t X & =  i[ H_{\rm sys}, X] + \mathcal{L}\dg[ L] X   + [ L\dg, X ]  \bin(t)   + \bin\dg(t) [ X, L], 
\end{align}
where $\mathcal{L}\dg[L]X = L\dg XL- \half \left( L\dg L X + X L\dg L  \right)$. 
The well-known relation between input and output field operators is
\begin{equation} \label{eq:inputoutput1}
\bout(t) = L + \bin(t).
\end{equation}
Here, the $\bin(t)$ operators satisfy $[\bin(t) ,\bin\dg(s) ]= \delta(t-s)$, and are given by
\begin{equation}
\bin(t) = \frac{1}{\sqrt{2\pi}}\int d\omega \, b_0(\omega)e^{-i\omega (t-t_0)},
\end{equation}
where  $b_0(\omega)$ is $b(\omega)$ taken at time $t_0$. By taking the limit $t_0 \rightarrow -\infty$ we obtain the field operators for the
scattering eigenstates, i.e.\ states which are frequency eigenstates at the asymptotic past, denoted by 
\begin{equation*}
\bin\dg(\omega)\ket{0} = \ket{\omega^+}.
\end{equation*}
It is clear that $\bin(\omega)$ and $\bin(t)$ are related by
\begin{equation} \label{eq:binFourier}
\bin(t) = \frac{1}{\sqrt{2\pi}}\int d\omega \, \bin (\omega)e^{-i\omega t}.
\end{equation}
Analogously, $\bout(t)$ satisfy $[\bout(t) ,\bout\dg(s) ]= \delta(t-s)$ and are given by 
\begin{equation}
\bout(t) = \frac{1}{\sqrt{2\pi}}\int d\omega \, b_f(\omega)e^{-i\omega (t-t_f)},
\end{equation}
where again $b_f(\omega)$ is $b(\omega)$ taken at time $t_f$, and we can take $t_f \rightarrow \infty$ to obtain the frequency eigenstates
in the asymptotic future:
\begin{equation*}
\bout\dg(\omega)\ket{0} = \ket{\omega^-}.
\end{equation*}
Operators $\bout(\omega)$ and $\bout(t)$ satisfy a relation analogous to \cref{eq:binFourier}. For a detailed account on the relationship between the input-output and scattering formalisms, see Refs.~\cite{GardColl85,FanKocaShen10,RoyWilsFirs16}. 

In what follows, we will need the natural extension of these formalisms to two chiral input-output modes, $\ain$ and $\bin$. In vector form, the resulting equations of motion for an arbitrary system operator $X$ and the input-output relations are
\begin{align}
\partial_t  X  =&\, i[ H_T, X] + \mathcal{L}\dg[ L_T] X \notag \\
	&+ [ L\dg_T, X ]  F_{\rm in}(t)  + F_{\rm in}\dg(t) [ X, L_T],  \label{eq:SLHchirala} \\
F_{\rm out}(t) =& L_T + F_{\rm in}(t),\label{eq:SLHchiralb}
\end{align}
where $L_T= (L_A,L_B)^T$, $F_{\rm in / out}(t) = (a_{{\rm in / out}}(t),b_{{\rm in / out}}(t))^T$, $H_T = H_A + H_B+ H_{AB}$, and $H_{AB}$ is an interaction between systems.

\subsection{SLH formalism and cascaded open quantum systems}\label{sec:SLH}

Following the success of input-output theory, a methodology was developed for driving a quantum system with the output field of another \cite{KoloSoko87,Carm93,Gard93}. This was recently elaborated into the so-called SLH formalism \cite{GougJame09}, which is a powerful mathematical tool-set, embodied in a few algebraic relations, that greatly simplifies the process of networking quantum systems. 

The SLH formalism assigns to each individual site in a network a triple $(S,L,H)$. The $L$ and $H$ parameters correspond exactly to those that appear in \crefrange{eq:fieldop1}{eq:inputoutput1}  [or \crefrange{eq:SLHchirala}{eq:SLHchiralb}], encoding the system-field coupling and the system's internal Hamiltonian, respectively. We will drop the $S$ parameter for brevity since it will be trivial throughout this paper. Suppose then that there are two quantum systems, with LH parameters $G_A = (L_A,H_A)$ and  $G_B = (L_B,H_B)$, which we wish to combine. There are two ways of doing this: the {\em concatenation product} and the {\em series product}.

\begin{figure}[t]
\includegraphics[width=0.65\columnwidth]{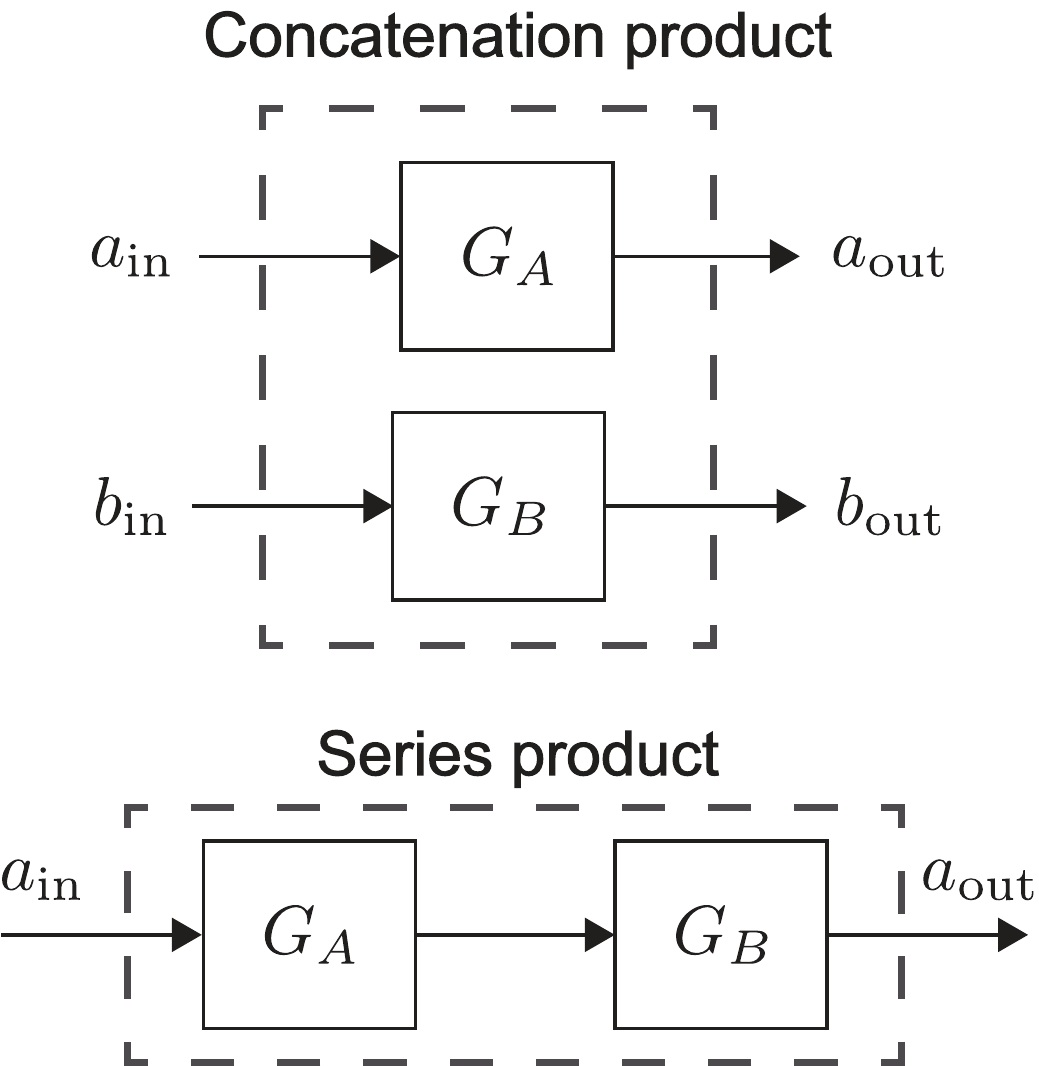}
\caption{Schematic representations of the concatenation product $G_A \boxplus G_A $ and series product 
$G_B \triangleleft G_A$. Both products are generalizations of the cascaded theory of open quantum systems~ \cite{GougJame09}.}\label{fig1}
\end{figure}

The {\em concatenation product} of  $G_A$ and  $G_B$, depicted in \Cref{fig1}, is a convenient way to describe the two independent systems as one larger system $G_T= (L_T, H_T)$, and is given by 
\begin{align}\label{eq:concatenation}
G_{T}=G_A \boxplus G_B = \left(\left[\begin{array}{c}
L_A \\ L_B \end{array}
\right],H_A+H_B \right).
\end{align}
Within this formalism, the concatenation product is the formal way of generalizing from \crefrange{eq:fieldop1}{eq:inputoutput1} to \crefrange{eq:SLHchirala}{eq:SLHchiralb}, or to higher numbers of input-output modes.
Now suppose you wish to feed the output of one system into the input of the other, as in \Cref{fig1}. This is the {\em series product}, and is given by 
\begin{align}
 G_T = & G_B \triangleleft G_A \notag \\
 = & \left(  L_A+L_B, H_A+H_B+\frac{1}{2i}(L_B^\dag L_A-L_A^\dag  L_B) \right).\label{eq:series}
\end{align}
After combining the two systems, the $L$ and $H$ parameters can be fed either into \Cref{eq:fieldop1} and \Cref{eq:inputoutput1} or into \Cref{eq:SLHchirala} and \Cref{eq:SLHchiralb}, as appropriate for the number of input/output modes, to obtain the relevant equations of motion.

These two relations may seem like trivial shortcuts, and indeed for small networks it is straightforward to perform these concatenations by hand. However, for larger or more complex networks this quickly becomes prohibitive, and the SLH formalism allows us to describe the entire network as a single large quantum system from which the input-output relations and equations of motion can be read off directly. Throughout this paper we will describe our networks of interaction sites simply by their SLH parameters. For a more in-depth review on this formalism, we direct the interested reader to \cite{GougJame09}.

\section{Single-site scattering}\label{sec:single}

Now we outline the steps required to determine the scattering matrix for one- and two-photon transport when 
a cross-Kerr interaction is present, i.e. a $\chi^{(3)}$ interaction. We consider several alternative physical realizations for the nonlinear interaction, and show they all have the same one- and two-photon S-matrices. Some of the results of this Section have previously been obtained in \cite{ViswGeaB15} using a different, but equivalent, formalism. However, it will be convenient to describe them here in detail, so as to outline the main mathematical steps used for our new results in subsequent sections.

\subsection{Cavity mediated Kerr interaction}
\begin{figure}[ht]
\includegraphics[width=0.7\columnwidth]{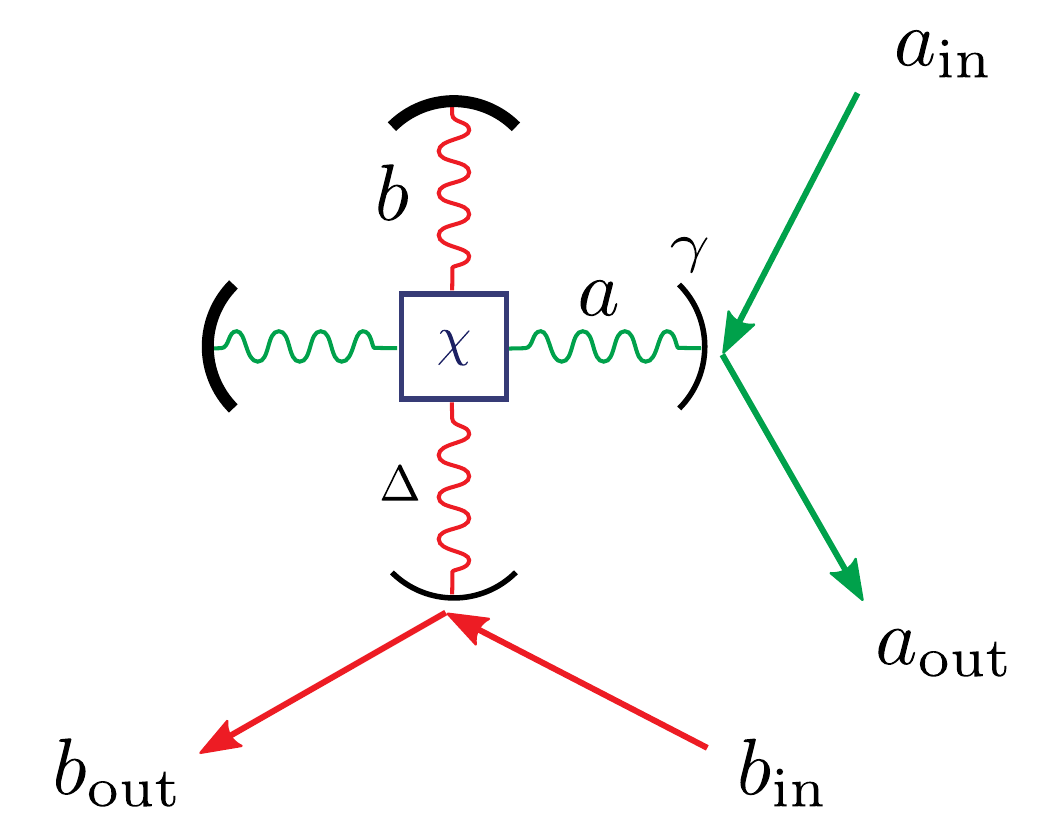}
\caption{(Color Online). An interaction between two input fields is mediated by two crossed cavities with a cross-Kerr interaction $H= \chi a\dg a b\dg b$. The input-output field in mode $a$ couples to the cavity internal energy $\Delta a\dg a$ and the input-output relation $\aout = \sqrt{\gamma}a +\ain$ (and similarly for mode $b$).}\label{fig2}
\end{figure}

The first system we wish to model is of two input fields that impinge on two cavities which interact via a cross-Kerr interaction, as illustrated in \Cref{fig2}. As usual, cavity operators obey $[a,a\dg]=[b,b\dg]=1$ and $[a,b\dg]=0$. The two cavities are identical, with internal Hamiltonians $\Delta a\dg a$ and $\Delta b\dg b$ \footnote{Although $\Delta$ is usually reserved for a detuning, rather than a cavity's resonant  frequency, we are here thinking of the cavity field as a harmonic oscillator, so $\Delta$ is the difference between its consecutive energy levels, just as in a subsequent section it will represent the energy difference between atomic levels.}, and the Kerr interaction is given by  $\chi a\dg a  b\dg b$. The cavities are coupled to the waveguides via operators $L_A=\sqrt{\gamma} a$ and $L_B=\sqrt{\gamma} b$. 

The main reason we have drawn this system as two separate cavities is in order to provide a natural way to separate the photons after the scattering interaction, which is essential for the concatenation scheme we have in mind.  Note, however, that the same goal could be accomplished with a single, one-sided cavity, provided the two photons have orthogonal polarizations: they could then be combined going in (and separated going out) at a polarizing beam splitter.

The total system $G_{T} = (L_T,H_T)$ is specified by the concatenation product (see \Cref{sec:SLH})
\begin{align}
G_{\rm sys} 
&=(\sqrt{\gamma} a, \Delta a\dg a +\chi a\dg a b\dg b )\boxplus (\sqrt{\gamma} b, \Delta b\dg b  ) \notag\\
&=\left(\left(\begin{array}{c}\sqrt{\gamma} a \\\sqrt{\gamma} b\end{array}\right), \Delta a\dg a +\Delta b\dg b +\chi a\dg a b\dg b \right).\label{eq:cavitySLH}
\end{align}
On the second line we have adopted the convention that the $a$ mode has the Kerr Hamiltonian. We can determine the equation of motion for the cavity operator using \cref{eq:SLHchirala} with $X=a$ and the LH parameters in \cref{eq:cavitySLH}
\begin{align}
 \partial_t a =&  - \left(\frac{\gamma}{2} + i \Delta \right) a  - i\chi  a b\dg b + \sqrt{\gamma}[ a\dg, a ] \ain(t) \nonumber \\
     =&  - \left(\frac{\gamma}{2} + i \Delta \right) a  - i\chi  a b\dg b - \sqrt{\gamma} \ain(t) . \label{eq:DE0sa}  
\end{align}
In going from the first to the second line we used $[ a\dg, a ] =-1$, a trivial simplification that will become important later on. Similarly,  the
equation of motion for the cavity operator $b$ is 
\begin{align}
 \partial_t b =&  -\left(\frac{\gamma}{2} + i \Delta \right) b  - i\chi  a\dg a  b - \sqrt{\gamma} \bin(t).
\end{align}
From Eqs.~(\ref{eq:SLHchiralb}) and (\ref{eq:cavitySLH}), the associated input--output relations are:
\begin{subequations}
\begin{align}
\aout(t) &= \sqrt{\gamma}\,a  + \ain(t) \\
\bout(t) &= \sqrt{\gamma}\,b  + \bin(t) .
\end{align}
\end{subequations}

\subsection{Atomic realizations of a Kerr interaction}

It has been argued that cross-Kerr interactions are only effective descriptions of interactions, as inevitably atoms must mediate the interaction~\cite{FanKockComb13}. For this reason we wish to consider here a couple of alternative realizations of the Kerr interaction involving one or two atoms.  

\begin{figure}[ht]
\includegraphics[width=\columnwidth]{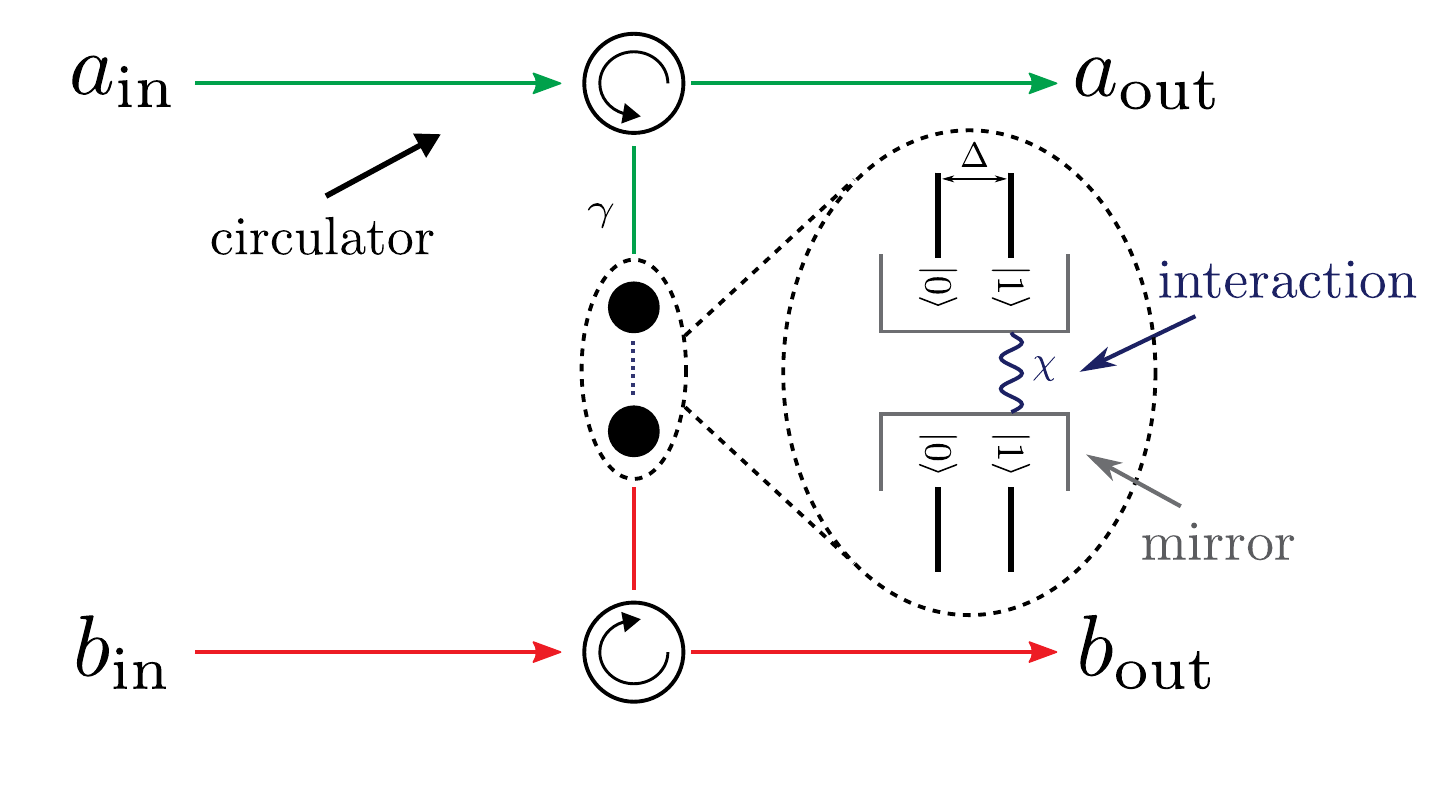}
\caption{(Color Online). An interaction between two input fields is mediated by two atoms that interact via $H=  \chi (\Id- A_{z})(\Id- B_{z})= \chi\op{1,1}{1,1}$. The mirrors ensure none of the field from mode $a$ leaks into mode $b$ and vice versa. The circulators ensure all of the input field arrives at the output. We show that, when at most one photon is present in each mode, the S-matrix for this case is identical to the cavity-mediated Kerr interaction of \cref{fig2}.}\label{fig3}
\end{figure}

We first consider two independent fields coupled to separate atoms that interact via a spin-spin interaction, as illustrated in \cref{fig3}. Each atom may be assumed to be near the closed end of a one-dimensional waveguide, yet they can interact through the mirror that separates them, via an interaction term of the form $\chi (\Id- A_{z}) (\Id- B_{z})$. 
The main reason to separate the atoms by a mirror is the same as discussed in the previous subsection; namely, we need a way to separate the output modes after the interaction so we can cascade the unit cell using circulators. Below we discuss an alternative polarization-based scheme which will likely be easier to implement.

The atomic lowering and raising operators, which couple the fields to the atoms, obey the standard commutation relations $[A_{-},A_{+}]= A_z$. The self Hamiltonians for atoms $A$ and $B$ are  $\frac{\Delta}{2} (\Id - A_{z})$ and $\frac{\Delta}{2} (\Id - B_{z})$, respectively, and the interaction between the atoms is given by $\chi (\Id- A_{z})(\Id- B_{z})$. The total system $G_{T} = (L_T,H_T)$ is specified by the concatenation product 
\begin{align}
G_{\rm sys} 
= & (\sqrt{\gamma} A_{-}, \frac{\Delta}{2} (\Id - A_{z}) + \chi (\Id- A_{z}) (\Id- B_{z}) ) \notag \\ 
   & \boxplus (\sqrt{\gamma}B_{-} , \frac{\Delta}{2} (\Id - B_{z})) \nonumber\\
= & \bigg( \left(\begin{array}{c}\sqrt{\gamma} A_{-} \\ \sqrt{\gamma} B_{-}\end{array}\right), \frac{\Delta}{2} (\Id - A_{z}) + \frac{\Delta}{2} (\Id - B_{z}) \notag \\ & + \chi (\Id- A_{z}) (\Id- B_{z}) \bigg).\label{unitCellSLHModel}
\end{align}
From this we determine the equation of motion for the atomic operator $A_{-}$ 
\begin{align}
    \partial_t A_{-} & = - \left(\frac{\gamma}{2} + i \Delta \right)A_{-} - i \chi A_{-}(\Id - B_{z}) - \sqrt{\gamma} A_z \opin{a}(t) \label{eq:DE1sa}.
\end{align}
Notice that this equation is almost identical to \cref{eq:DE0sa}. The only differences are the $\sqrt{\gamma}  A_z \opin{a}$ and $\chi A_{-}(\Id - B_{z})$ terms, which for the cavity case read $\sqrt{\gamma} \ain$ and $i \chi  a b\dg b$ respectively. These differences originate in slightly different commutation relations obeyed by cavity and atomic operators but, as we shall see in the next two sections, they have no effect on the subset of single- and two-photon transport that we consider. If one was to consider, however, two- or multi-photon transport where any input mode has more than one photon, then the atomic medium would begin to saturate and the differences between the cavity- and atom-mediated nonlinearities would likely manifest.

Similarly the equation of motion for the operator $B_{-}$  is
\begin{align}
    \partial_t B_{-} & = - \left(\frac{\gamma}{2} + i \Delta \right)B_{-} - i \chi (\Id -A_{z}) B_{-} - \sqrt{\gamma} B_z \opin{b}(t), \label{eq:DE1sb}
\end{align}    
and the input-output relations are
\begin{align}
    \opout{a}(t) & = \opin{a}(t) + \sqrt{\gamma} A_{-} \label{eq:DE1sc}\\
    \opout{b}(t) & = \opin{b}(t)+ \sqrt{\gamma} B_{-} \label{eq:DE1sd}.
\end{align}

\begin{figure}[ht]
\includegraphics[width=0.8\columnwidth]{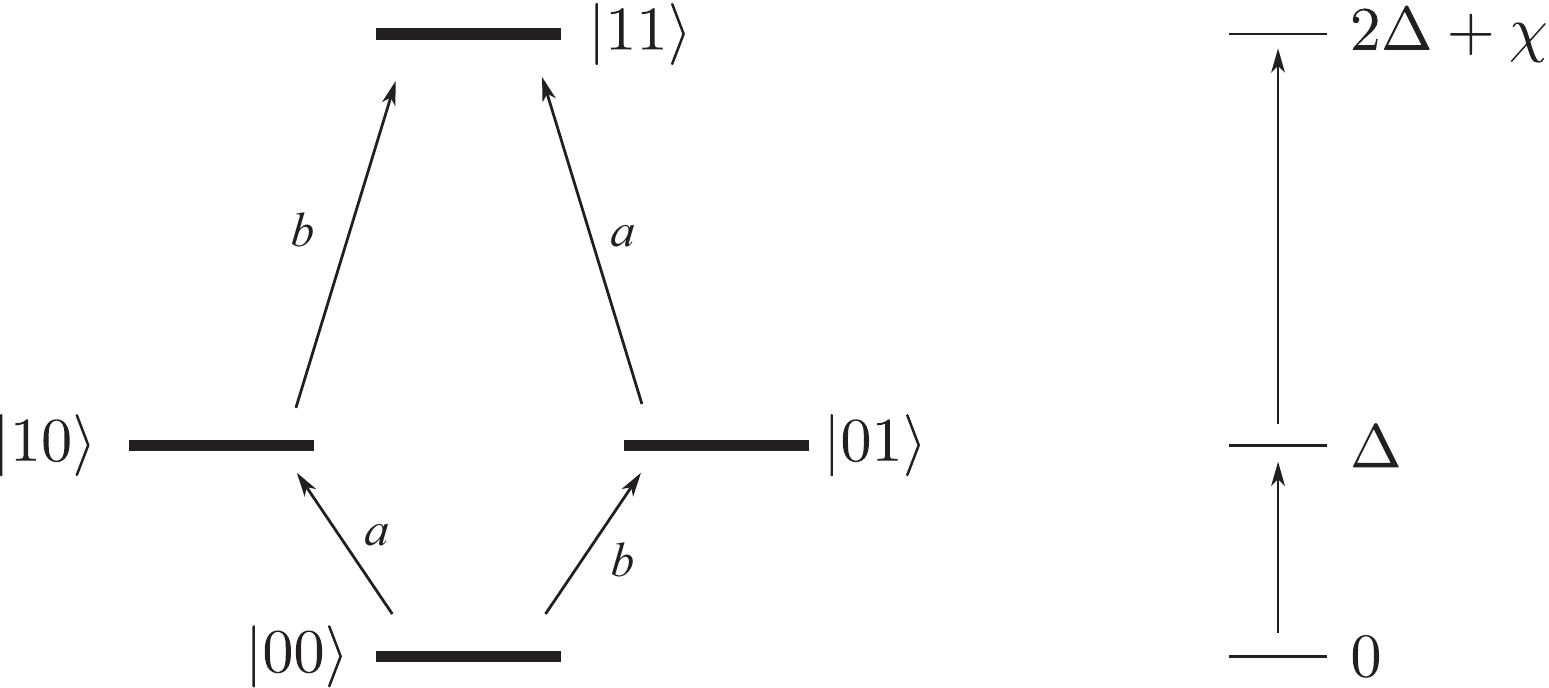}
\caption{The level structure of two coupled two-level atoms. In the $\chi\rightarrow \infty$ limit, the population of the $\ket{1, 1}$ atomic state is suppressed, and the remaining three atomic states can be identified with the states of a single three-level atom in a ``V'' configuration. }\label{fig4}
\end{figure}

Many physically different arrangements lead to the same interaction, i.e. \cref{unitCellSLHModel}, and subsequent equations of motion and input-output relations. For example, it is straightforward to see that these equations would equally well describe a single four-level atom, in a one-sided waveguide or cavity (the latter only in the ``fast-cavity'' regime), with the level scheme shown in \cref{fig4}, if we assume that only the $a$ photons can excite the $\ket{00}\to\ket{10}$ and $\ket{01}\to\ket{11}$ transitions, and likewise only the $b$ photons can excite the $\ket{00}\to\ket{01}$ and $\ket{10}\to\ket{11}$ transitions.  One way this might be arranged would be for the photons to have orthogonal circular polarizations, the levels $\ket{00}$ and $\ket{11}$ to have magnetic quantum number $m=0$, and the levels $\ket{10}$ and $\ket{01}$ to have $m=\pm 1$.

In the level scheme illustrated in \cref{fig4}, the role of the ``coupling'' $\chi$ is played by the detuning of the uppermost level from exact two-photon resonance.  As we shall see below, only when $\chi$ is equal to zero is the outgoing two-photon state unentangled (assuming it was unentangled going in), since in that case the system is equivalent to just two non-interacting two-level atoms.  For very large $\chi$, on the other hand, one may in effect ignore the uppermost level altogether, and then the system is equivalent to the three-level, ``V'' configuration discussed by Koshino \cite{Kosh09} and Chudzicki \etal~\cite{ChudChuaShap13}.

We now turn our attention to the asymptotic output states of the interaction (\ref{unitCellSLHModel}) when (i) a single photon is present either in mode $\ain$ or mode $\bin$, and (ii) a single photon is present in both modes $\ain$ and $\bin$. At relevant points in the derivation we will note the similarities and differences between the calculations involving cavity operators $a$ and $b$ and atomic operators $A_{-}$ and $B_{-}$.

\subsection{Single-photon S-matrix}\label{sec:1site1photon}

Our method and presentation closely follow those of Ref. \cite{FanKocaShen10}. We will solve for the S-matrix when there is an input photon in mode $\ain$, by symmetry the solution for single-photon transport in mode $\bin$ is identical. The elements of the S-matrix can be specified in the frequency domain by
\begin{equation}\label{eq:1scatmatrix0}
S_{\omega_a  ,\nu_a } = \ip{\omega_a ^{-}}{\nu_a^{+}},
\end{equation}
which we can express as \cite{FanKocaShen10}:
\begin{equation}\label{eq:1scatmatrix}
S_{\omega_a  ,\nu_a } = \frac{1}{\sqrt{2\pi}} \int dt \bra{0} \opout{a} (t) \ket{\nu_a^{+}} e^{i \omega_a  t}.
\end{equation}
By using the input-output relation, i.e.\ \cref{eq:DE1sc}, this becomes
\begin{equation} \label{eq:1p1sscattmain}
\braket{\omega_a ^{-}|\nu_a^{+}} = \frac{1}{\sqrt{2\pi}} \int dt \bra{0} \opin{a}(t) + \sqrt{\gamma} A_{-} \ket{\nu_a^{+}} e^{i \omega_a  t},
\end{equation}
The first term simply reduces to $\delta(\omega_a - \nu_a)$, due to the delta commutation relations of the input operators. The second term can be obtained from \cref{eq:DE1sa} by sandwiching it between $\bra{0}$ and $\ket{\nu_a^{+}}$, leading to the differential equation
\begin{align}
\partial_t \bra{0} A_{-} \ket{\nu_a^{+}} = & - \left(\frac{\gamma}{2} + i \Delta \right) \bra{0} A_{-} \ket{\nu_a^{+}} \notag\\ 
& - \sqrt{\gamma} \bra{0} A_{z} \opin{a}(t) \ket{\nu_a^{+}},\label{eq:1p1sDEmatrix1}
\end{align}
where the matrix element coming from the interaction term in \cref{eq:DE1sa}, proportional to $\chi$, is zero since the second mode is in vacuum. As the atom is in the ground state, it holds that $ \bra{0}A_z=\bra{0}$, and the solution to this equation is identical to what we would have obtained by starting from \cref{eq:DE0sa} instead, with an initially empty cavity. By solving this differential equation, as detailed in \Cref{apx:singlesystem1p}, we obtain the single-photon scattering matrix:
\begin{equation} \label{eq:1p1sscatt2}
S_{\omega_a  ,\nu_a } =\braket{\omega_a ^{-}|\nu_a^{+}} = - \frac{\conj{\Gamma}(\omega_a )}{\Gamma(\omega_a )} \delta(\omega_a  - \nu_a).
\end{equation}
where we defined the shorthand
\begin{align} 
\Gamma(\omega) := \frac{\gamma}{2} + i (\Delta-\omega).
\end{align}
\Cref{eq:1p1sscatt2} has been derived many times in the literature, e.g.~\cite{YurkDenk84,CollGard84}.

Now consider an input photon in some frequency-domain wave packet $\ket{1_\xi}:=\int d\nu \, \xi(\nu) \opin{a^{\dagger}}(\nu) \ket{0}$. \Cref{eq:1p1sscatt2} can be used to obtain the output wave packet by inserting a resolution of the identity (on the output modes):

\begin{align*}
\ket{1_{\xi^{'}}} & = \int d\nu \, d\omega \, S_{\omega ,\nu } \xi(\nu) \opout{a^{\dagger}}(\omega) \ket{0} \\
& = \int d\omega \, \xi^{'}(\omega) \opout{a^{\dagger}}(\omega) \ket{0},
\end{align*}
where $\xi^{'}(\omega)\! := \int d\nu \,  S_{\omega ,\nu } \xi(\nu)  $ and similarly for two-photon transport.

\subsection{Two-photon S-matrix}\label{sec:1site2photon}
The particular kind of two-photon transport we wish to consider involves a single input photon in mode $\ain$ and a single input photon in mode $\bin$. The S-matrix for this problem is
\begin{align*}
S_{\omega_a  \omega_b,\nu_a \nu_b} & = \braket{\omega_a ^{-} \omega_b^{-}|\nu_a^+ \nu_b^+} \\ &= \bra{0_a,0_b} \opout{a}(\omega_a )\opout{b}(\omega_b)\ket{\nu_a^+ \nu_b^+}.
\end{align*}
To determine the S-matrix we use the input-output relations given in \cref{eq:DE1sd} and the single-photon results from \Cref{sec:1site1photon}. Doing so gives
\begin{align}
\braket{\omega_a ^{-} \omega_b^{-}|\nu_a^+ \nu_b^+} = & - \frac{\conj{\Gamma}(\omega_a )}{\Gamma(\omega_a )} \bigg( \delta(\omega_a  - \nu_a) \delta(\omega_b - \nu_b) \notag \\
& + \sqrt{\frac{\gamma}{2\pi}} \int dt \,e^{i \omega_b t} \bra{\omega_a ^+} B_{-} \ket{\nu_a^+ \nu_b^+} \bigg).  \label{eq:2p1sscatt}
\end{align}
To solve for the matrix element $\bra{\omega_a ^+} B_{-} \ket{\nu_a^+ \nu_b^+}$ we sandwich \cref{eq:DE1sb} between $\bra{\omega_a ^+}$ and $\ket{\nu_a^+ \nu_b^+}$, as before, to obtain
\begin{align}
\partial_t \bra{\omega_a ^+} B_{-} \ket{\nu_a^{+} \nu_b^+} =& - \left(\frac{\gamma}{2} + i \Delta \right) \bra{\omega_a ^+} B_{-} \ket{\nu_a^{+} \nu_b^+} \notag \\
&- i \chi \bra{\omega_a ^+} (\Id-A_z) B_{-} \ket{\nu_a^{+} \nu_b^+} \notag \\
&- \sqrt{\gamma} \bra{\omega_a ^+} B_{z} \opin{b}(t) \ket{\nu_a^{+} \nu_b^+}.  \label{eq:2p1sDEmatrix3}
\end{align}
As before we have $\bra{0}B_{z}=\bra{0}$ and furthermore $(\Id-A_z)$ corresponds to the number operator in the single-photon sector for the cavity in mode $a$. Consequently the solution to this equation is identical to the cavity-mediated case. It is not obvious how to solve this equation for $ \bra{\omega_a ^+} B_{-} \ket{\nu_a^{+} \nu_b^+}$ due to the interaction with the photon in the other mode, i.e. the $\bra{\omega_a ^+} (\Id-A_z) B_{-} \ket{\nu_a^{+} \nu_b^+} $ term. Using a few operator identities, as detailed in the \Cref{apx:singlesystem2p},  it can be reduced to a more manageable form:
\begin{align} 
\partial_t& \bra{\omega_a ^+} B_{-} \ket{\nu_a^{+} \nu_b^+}\nonumber\\
=&  - \left(\frac{\gamma}{2} + i \Delta \right) \bra{\omega_a ^+} B_{-} \ket{\nu_a^{+} \nu_b^+} \notag \\
& - i \frac{\chi \gamma}{\pi} \frac{1}{\conj{\Gamma}(\omega_a )} \int dp_a\frac{e^{-i (p_a-\omega_a )t}}{\Gamma(p_a)} \bra{p_a^+} B_{-} \ket{\nu_a^{+} \nu_b^+} \notag \\
& - \sqrt{\frac{\gamma}{2\pi}} e^{- i \nu_b t} \delta(\omega_a  - \nu_a). \label{eq:2p1sDEmatrix4}
\end{align}
Since \cref{eq:2p1sscatt} actually requires the Fourier transform of $\bra{\omega_a ^+} B_{-} \ket{\nu_a^{+} \nu_b^+}$, it is convenient to move \cref{eq:2p1sDEmatrix4} into the Fourier domain, reducing it to an integral equation. The solution, detailed in \Cref{apx:singlesystem2p}, leads to the S-matrix:
\begin{align}
S_{\omega_a  \omega_b,\nu_a \nu_b}=& \braket{\omega_a ^{-} \omega_b^{-}|\nu_a^+ \nu_b^+} \nonumber\\
=& \braket{\omega_a ^{-}|\nu_a^{+}}\braket{\omega_b^{-}|\nu_b^{+}} \nonumber\\
& - i \frac{\chi \gamma^2}{\pi} \left( 1+\frac{2 i \chi}{\Gamma(\omega_a )+\Gamma(\omega_b)} \right)^{-1}\nonumber\\
& \times \frac{\delta(\omega_a  + \omega_b - \nu_a - \nu_b)}{\Gamma(\nu_b)\Gamma(\nu_a) \Gamma(\omega_b) \Gamma(\omega_a )}, \label{eq:2p1sfinal}
\end{align}
where $\braket{\omega_a ^{-}|\nu_a^{+}}\braket{\omega_b^{-}|\nu_b^{+}} $ are the contributions from single-photon transport, i.e. \cref{eq:1p1sscatt2}. The term proportional to $\chi$ is, of course, due to the Kerr interaction. The delta function $\delta(\omega_a  + \omega_b - \nu_a - \nu_b)$ arises from energy conservation, and is generally responsible for spectral entanglement in the output photon wave packets.  
 
As mentioned previously, a result equivalent to \cref{eq:2p1sfinal} was derived by a direct calculation of the output wave function in Ref.~\cite{ViswGeaB15}, although the actual result presented there [Eq.~(41) of Ref.~\cite{ViswGeaB15}] has been simplified under the assumption that the joint spectrum of the incoming photons is symmetric with respect to the exchange of $\nu_a$ and $\nu_b$.

\subsection{Reduction to three-level atom S-matrix}
One can also check that in the limit $\chi\rightarrow \infty$, the two-photon S-matrix [\cref{eq:2p1sfinal}] reduces to  
\begin{align}
S_{\omega_a  \omega_b,\nu_a \nu_b} = &  \braket{\omega_a ^{-}|\nu_a^{+}}\braket{\omega_b^{-}|\nu_b^{+}} - \frac{ \gamma^2}{2\pi} \left[\frac{1}{\Gamma(\omega_a )}+\frac{1}{\Gamma(\omega_b)} \right] \notag \\
& \times \frac{1}{\Gamma(\nu_b)\Gamma(\nu_a) }\delta(\omega_a  + \omega_b - \nu_a - \nu_b), \label{eq:TRLfinal}
\end{align}
which is identical to the two-photon S-matrix for a three-level atom in a ``V" configuration as derived by Koshino \cite{Kosh09} and Chudzicki \etal~\cite{ChudChuaShap13}, up to a redefinition of the coupling constant $\gamma$. Hence our model contains that system as a limit, as explained in connection with \cref{fig4} above.

\section{Two-site scattering with co-propagating photons}\label{sec:2coprop}

In this section we will take the single-site Kerr interaction and feed its output into the input of another similar interaction, with both photons propagating in the same direction. A possible physical realization is illustrated in \Cref{fig4}, which achieves directionality using optical isolators or circulators.

\begin{figure}[ht]
\includegraphics[width=0.7\columnwidth]{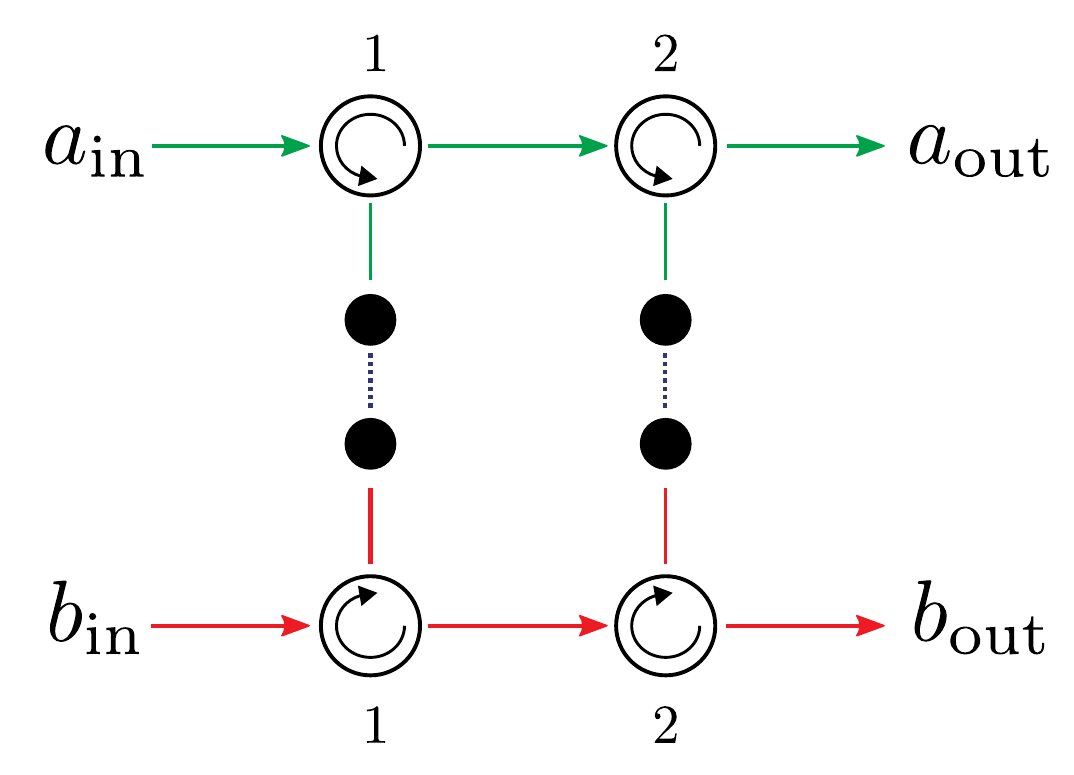}
\caption{(Color online). A two-site Kerr interaction with co-propagating photons.}\label{fig5}
\end{figure}

To derive the system operator differential equations, input-output relations, and subsequently the S-matrix, we use the SLH formalism described in \Cref{sec:SLH}.

\subsection{Cascading and differential equations}
The $i$th site is specified by the interaction \cref{unitCellSLHModel} with parameters $\chi_i$,$\gamma_i$, and $\Delta_i$.
Then the cascading of two site interaction where both photons are propagating in the same direction has an SLH model
\begin{align}
G_{\rm sys} 
&= G_{\rm A}\boxplus G_{\rm B} \nonumber\\
&= \left (L_{\rm A} , H_{\rm A} \right)\boxplus \left ( L_{\rm B} , H_{\rm B} \right) \nonumber\\
&= \left ( G_{\rm A}^{(2)}\lhd G_{\rm A}^{(1)} \right )\boxplus \left ( G_{\rm B}^{(2)}\lhd G_{\rm B}^{(1)} \right ), \label{eq:SLHco}
\end{align}
where
\begin{align}
L_{\rm A}=&\sqrt{\gamma_1}  \opmi{A}{1}+\sqrt{\gamma_2}  \opmi{A}{2},\\
H_{\rm A}=& \frac{\Delta_1}{2} (\Id - \opz{A}{1}) +\chi_1 (\Id-  \opz{A}{1}) (\Id-  \opz{B}{1}) \nonumber\\
&+ \frac{\Delta_2}{2} (\Id - \opz{A}{2}) +\chi_2 (\Id-  \opz{A}{2}) (\Id-  \opz{B}{2}) \nonumber \\
& + \frac{\sqrt{\gamma_1 \gamma_2}}{2i}\left (\oppl{A}{2} \opmi{A}{1} - \oppl{A}{1} \opmi{A}{2} \right ),
\end{align}
and
\begin{align}
L_{\rm B}=&\sqrt{\gamma_1}  \opmi{B}{1}+\sqrt{\gamma_2}  \opmi{B}{2},\\
H_{\rm B}=&\frac{\Delta_1}{2}(\Id- \opz{B}{1})+\frac{\Delta_2}{2}(\Id- \opz{B}{2}) \nonumber \\
& + \frac{\sqrt{\gamma_1 \gamma_2}}{2i}\left (\oppl{B}{2} \opmi{B}{1} - \oppl{B}{1} \opmi{B}{2}  \right ). \label{eq:HBco}
\end{align}
Once more, we have used the convention that the Kerr interaction is included in $H_{\rm A}$.

After substituting the SLH parameters and going through the commutator algebra we have the following differential equations
\begin{widetext}
\begin{subequations} \label{eq:DE2scop}
    \begin{align}
    \partial_t \opmi{A}{1} & = - \left(\frac{\gamma_1}{2}+i \Delta_1 \right) \opmi{A}{1} - i \chi_1 \opmi{A}{1} (\Id-\opz{B}{1}) - \sqrt{\gamma_1}\opz{A}{1} \opin{a}(t),  \label{eq:DE2scopa} \\
    \partial_t \opmi{B}{1} & = - \left(\frac{\gamma_1}{2}+i \Delta_1 \right) \opmi{B}{1} - i \chi_1 (\Id-\opz{A}{1})\opmi{B}{1} - \sqrt{\gamma_1}\opz{B}{1} \opin{b}(t) , \label{eq:DE2scopb} \\
    \partial_t \opmi{A}{2} & = - \left(\frac{\gamma_2}{2}+i \Delta_2 \right) \opmi{A}{2} - i \chi_2 \opmi{A}{2}(\Id-\opz{B}{2}) - \sqrt{\gamma_1 \gamma_2}\opz{A}{2} \opmi{A}{1} -\sqrt{\gamma_2}\opz{A}{2} \opin{a}(t),  \label{eq:DE2scopc} \\
    \partial_t \opmi{B}{2} & = - \left(\frac{\gamma_2}{2}+i \Delta_2 \right) \opmi{B}{2} - i \chi_2 (\Id-\opz{A}{2})\opmi{B}{2} - \sqrt{\gamma_1 \gamma_2}\opz{B}{2} \opmi{B}{1} -\sqrt{\gamma_2} \opz{B}{2} \opin{b}(t) , \label{eq:DE2scopd} 
    \end{align}
\end{subequations} 
\end{widetext}    
and the input-output relations
\begin{subequations}
\begin{align}
\opout{a}(t) & = \opin{a}(t) + \sqrt{\gamma_1} \opmi{A}{1} +\sqrt{\gamma_2} \opmi{A}{2},  \label{eq:DE2scope} \\
\opout{b}(t) & = \opin{b}(t) + \sqrt{\gamma_1} \opmi{B}{1} +\sqrt{\gamma_2} \opmi{B}{2}.  \label{eq:DE2scopf} 
 \end{align}
\end{subequations}
The differential equations and input-output relations above are almost identical to the cascaded cavity-mediated Kerr interaction. As before, the differences between the atomic Kerr interaction and the cavity case are the $\sqrt{\gamma_i}  \opz{A}{i} \opin{a}$ and $\chi_i A_{-}(\Id - \opz{B}{i})$ etc. terms, which for the cavity case read $\sqrt{\gamma_i} \ain$ and $i \chi_i  a_i b_i\dg b_i$ respectively. Nevertheless, these differences do not affect the solution to the single- and two-photon transport that we consider, and this remains true in the counter-propagating case.

\subsection{Single-photon S-matrix}

Structurally, the S-matrix for single-photon transport looks similar to \cref{eq:1scatmatrix}. However, this time we must use the two-site input-output relation, \cref{eq:DE2scope}, to obtain:
\begin{align} 
\braket{\omega_a ^{-}|\nu_a^{+}} = & \delta(\omega_a  - \nu_a) \notag \\ 
& +  \sqrt{\frac{\gamma_1}{2\pi}} \int dt \bra{0} \opmi{A}{1} \ket{\nu_a^{+}} e^{i \omega_a  t} \notag \\
& + \sqrt{\frac{\gamma_2}{2\pi}} \int dt \bra{0} \opmi{A}{2} \ket{\nu_a^{+}} e^{i \omega_a  t}. \label{eq:1p2scopscatt}
\end{align}
The differential equation for $\bra{0} \opmi{A}{1} \ket{\nu_a^{+}}$ is essentially the same as \cref{eq:1p1sDEmatrix1}. The differential equation for $\bra{0} \opmi{A}{2} \ket{\nu_a^{+}}$ can be obtained from \cref{eq:DE2scopc}:
\begin{align}
\partial_t \bra{0} \opmi{A}{2} \ket{\nu_a^{+}} = & - \left(\frac{\gamma_2}{2} + i \Delta_2 \right) \bra{0} \opmi{A}{2} \ket{\nu_a^{+}} \notag \\
& - \sqrt{\gamma_1 \gamma_2} \bra{0} \opmi{A}{1} \ket{\nu_a^{+}} \notag \\
& - \sqrt{\gamma_2} \bra{0} \opin{a}(t) \ket{\nu_a^{+}}. \label{eq:1p2sDEmatrix1b}
\end{align}
Notice that \cref{eq:1p2sDEmatrix1b} depends on $\bra{0} \opmi{A}{1} \ket{\nu_a^{+}}$, so we must solve the equation for the latter first, then use that  result in the former. The steps for this solution are essentially the same as in the single-site case, and are carried out in \Cref{apx:twosystemcop1p}, leading to the final expression
\begin{align} \label{eq:1p2scopscatt2}
\braket{\omega_a ^{-}|\nu_a^{+}}  = \frac{\conj{\Gamma_2}(\omega_a ) \conj{\Gamma_1}(\omega_a )}{\Gamma_2(\omega_a )\Gamma_1(\omega_a )} \delta(\omega_a  - \nu_a), 
\end{align}
where we defined 
\begin{align} 
\Gamma_i(\omega) := \frac{\gamma_i}{2} + i (\Delta_i-\omega).
\end{align}
\Cref{eq:1p2scopscatt2} is as expected---the single photon simply passes through both systems, acquiring the cumulative effect of coupling with both atoms.

\subsection{Two-photon S-matrix}
The fundamental difference between the single-site and two-site single-photon transports was the additional term arising from the cascaded input-output relations.
This is also the case for two-site two-photon transport. Our starting point is similar to \cref{eq:2p1sscatt}, for the single-site case, we use  \cref{eq:DE2scopf} to obtain
\begin{align} 
\braket{\omega_a ^{-} \omega_b^{-}|\nu_a^+ \nu_b^+} = & \frac{\conj{\Gamma_1}(\omega_a )\conj{\Gamma_2}(\omega_a )}{\Gamma_1(\omega_a )\Gamma_2(\omega_a )} \notag \\
& \times \bigg( \delta(\omega_a  - \nu_a) \delta(\omega_b - \nu_b) \notag \\
& + \sqrt{\frac{\gamma_1}{2\pi}} \int dt e^{i \omega_b t} \bra{\omega_a ^+} \opmi{B}{1} \ket{\nu_a^+ \nu_b^+} \notag \\
& +\sqrt{\frac{\gamma_2}{2\pi}} \int dt e^{i \omega_b t} \bra{\omega_a ^+} \opmi{B}{2} \ket{\nu_a^+ \nu_b^+} \bigg). \label{eq:2p2scopscatt}
\end{align}
The equation for $\bra{\omega_a ^+} \opmi{B}{1} \ket{\nu_a^+ \nu_b^+}$ is the same as in the single-site case [cf.\ \cref{eq:2p1sDEmatrix3}], thus we simply use that solution. Now we take the $\bra{\omega_a ^+} \cdot  \ket{\nu_a^+ \nu_b^+}$ matrix element of \cref{eq:DE2scopd} and, making similar manipulations as those leading up to \cref{eq:2p1sDEmatrix4}, we arrive at
\begin{widetext}
\begin{align}
    \partial_t \bra{\omega_a ^+} \opmi{B}{2} \ket{\nu_a^{+} \nu_b^+} = & - \left(\frac{\gamma_2}{2} + i \Delta_2 \right) \bra{\omega_a ^+} \opmi{B}{2} \ket{\nu_a^{+} \nu_b^+} - \sqrt{\gamma_1 \gamma_2} \bra{\omega_a ^+} \opmi{B}{1} \ket{\nu_a^{+} \nu_b^+}  - \sqrt{\frac{\gamma_2}{2\pi}} e^{- i \nu_b t} \delta(\omega_a  - \nu_a)  \notag \\
    & - i \frac{\chi_2 \gamma_2}{\pi} \frac{\Gamma_1(\omega_a)}{\conj{\Gamma_1}(\omega_a )\conj{\Gamma_2}(\omega_a )}  \int dp_a \frac{\conj{\Gamma}_1(p_a)}{\Gamma_1(p_a) \Gamma_2(p_a)} e^{-i (p_a-\omega_a )t} \bra{p_a^+} \opmi{B}{2} \ket{\nu_a^{+} \nu_b^+}, \label{eq:2p2scopDEmatrix4b}
\end{align}     
\end{widetext}
As in the single-photon case, we have a system of one standalone equation (for $\bra{\omega_a ^+} \opmi{B}{1} \ket{\nu_a^{+} \nu_b^+}$) and one dependent equation (for $\bra{\omega_a ^+} \opmi{B}{2} \ket{\nu_a^{+} \nu_b^+}$). By solving both in the correct order and in the Fourier domain, as carried out in \Cref{apx:twosystemcop2p}, we obtain the final result
\begin{widetext}
\begin{align}
\braket{\omega_a ^{-} \omega_b^{-}|\nu_a^+ \nu_b^+}  = & \braket{\omega_a ^{-}|\nu_a^{+}}\braket{\omega_b^{-}|\nu_b^{+}}
-  \frac{\delta(\omega_a  + \omega_b - \nu_a - \nu_b)}{\pi} \notag \\ 
& \times \bigg[  i \left( \frac{\conj{\Gamma}_2(\omega_a) \conj{\Gamma}_2(\omega_b)}{\Gamma_2(\omega_a) \Gamma_2(\omega_b)} \right) \left( 1 + \frac{2 i \chi_1}{\Gamma_1(\omega_a) + \Gamma_1(\omega_b)} \right)^{-1} \frac{\chi_1 \gamma_1^2}{\Gamma_1(\nu_b)\Gamma_1(\nu_a) \Gamma_1(\omega_b) \Gamma_1(\omega_a )} \notag \\
& + i \left( \frac{\conj{\Gamma}_1(\nu_a) \conj{\Gamma}_1(\nu_b)}{\Gamma_1(\nu_a) \Gamma_1(\nu_b)} \right) \left( 1 + \frac{2 i \chi_2}{\Gamma_2(\omega_a) + \Gamma_2(\omega_b)} \right)^{-1} \frac{\chi_2 \gamma_2^2}{\Gamma_2(\nu_b)\Gamma_2(\nu_a) \Gamma_2(\omega_b) \Gamma_2(\omega_a )} \notag \\
& + \left( 1 + \frac{2 i \chi_1}{\Gamma_1(\omega_a) + \Gamma_1(\omega_b)} \right)^{-1} \left( 1 + \frac{2 i \chi_2}{\Gamma_2(\omega_a) + \Gamma_2(\omega_b)} \right)^{-1} \frac{4 \chi_1 \chi_2 \gamma_1^2 \gamma_2^2}{\Gamma_1(\nu_b)\Gamma_1(\nu_a) \Gamma_2(\omega_b) \Gamma_2(\omega_a )} \notag \\
& \times \frac{1}{(\Gamma_1(\omega_a) + \Gamma_1(\omega_b))(\Gamma_1(\omega_a) + \Gamma_2(\omega_b))(\Gamma_2(\omega_a) + \Gamma_2(\omega_b))} \bigg].
 \label{eq:2p2scopfinal}
\end{align}     
\end{widetext}

As unwieldy as this expression is, there is a clear physical interpretation behind it. The scattering matrix is the sum of four scattering channels: 
\begin{itemize}
\item[(I)] The single-photon term, where the photons don't interact at all. This corresponds to the first term in the sum, obtained by setting $\chi_1 = \chi_2 = 0$;
\item[(II)] The term where the photons interact only at the first atom, obtained by setting $\chi_2 = 0$. This amplitude can be understood as the product of the interaction term induced by $\chi_1$ and 
a \emph{single-photon} phase factor picked up by both photons at the second interaction site;
\item[(III)] Same as (II), but with the exchange $1 \leftrightarrow 2$, corresponding to the interaction happening only at the second site; and
\item[(IV)] The last term, proportional to $\chi_1 \chi_2$, which encodes the channel where the photons interact at the first site, and then again at the second site.
\end{itemize}

\section{Two-site scattering with counter-propagating photons}\label{sec:2counter}

\begin{figure}[b]
\includegraphics[width=0.7\columnwidth]{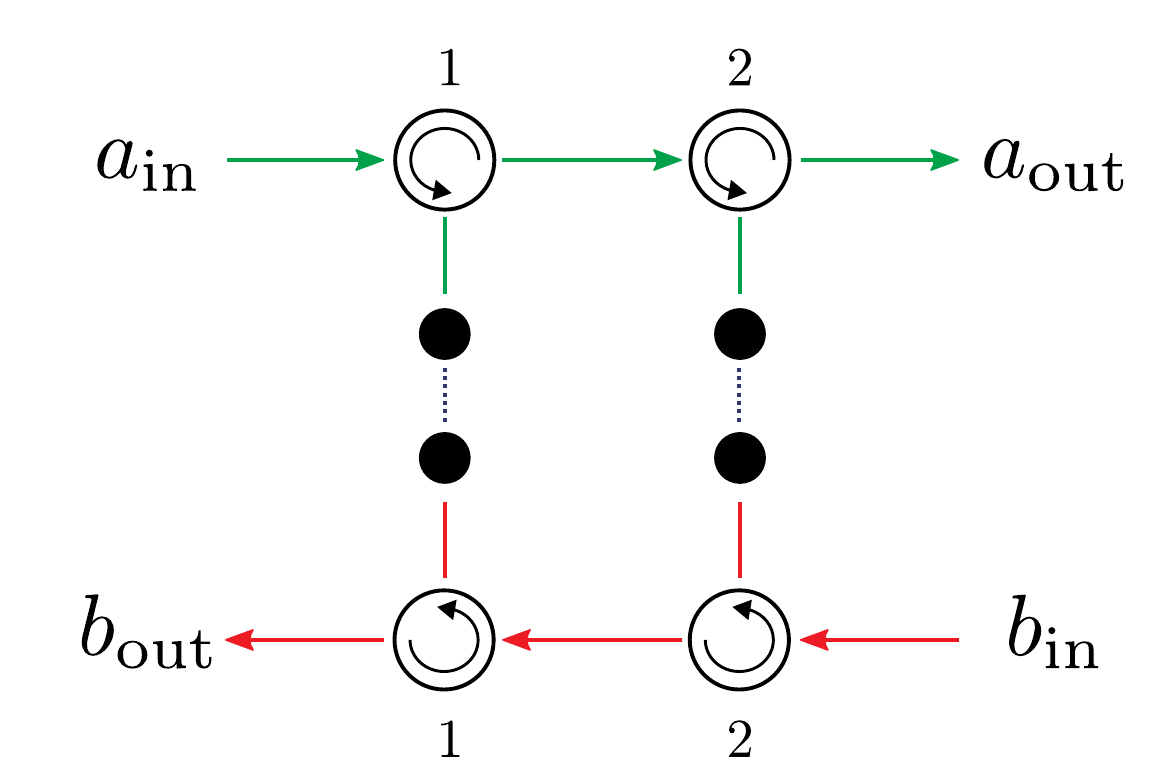}
\caption{(Color online) A two-site Kerr interaction with counter-propagating photons.}\label{fig6}
\end{figure}

In this section we will take the single-site Kerr interaction and feed its output into the input of another similar interaction, however now the photons in mode $\ain$ and $\bin$ will be propagating in opposite directions. A possible physical realization is illustrated in \Cref{fig6}.

Our starting point is, once more, the SLH formalism described in \Cref{sec:inputoutput}. 

\subsection{Cascading and differential equations}
The $i$th site is specified by the interaction \cref{unitCellSLHModel} with parameters $\chi_i$,$\gamma_i$, and $\Delta_i$.
Then, the cascading of two interaction sites where the photons are counter-propagating is modelled in the SLH formalism as follows
\begin{align}
G_{\rm sys}  
&= G_{\rm A}\boxplus G_{\rm B} \notag \\
&= \left (L_{\rm A} , H_{\rm A} \right)\boxplus \left ( L_{\rm B} , H_{\rm B} \right) \notag \\
&= \left ( G_{\rm A}^{(2)}\lhd G_{\rm A}^{(1)} \right )\boxplus \left ( G_{\rm B}^{(1)}\lhd G_{\rm B}^{(2)} \right ),
\end{align}
where
\begin{align}
L_{\rm A}=&\sqrt{\gamma_1}  \opmi{A}{1}+\sqrt{\gamma_2}  \opmi{A}{2},\\
H_{\rm A}=& \frac{\Delta_1}{2} (\Id - \opz{A}{1})+\chi_1 (\Id-  \opz{A}{1}) (\Id-  \opz{B}{1}) \nonumber\\
&+ \frac{\Delta_2}{2} (\Id - \opz{A}{2}) +\chi_2 (\Id-  \opz{A}{2}) (\Id-  \opz{B}{2}) \nonumber \\
&+ \frac{\sqrt{\gamma_1 \gamma_2}}{2i}\left ( \oppl{A}{2} \opmi{A}{1} - \oppl{A}{1}  \opmi{A}{2}  \right ),
\end{align}
and
\begin{align}
L_{\rm B}=&\sqrt{\gamma_1}  \opmi{B}{1}+\sqrt{\gamma_2}  \opmi{B}{2},\\
H_{\rm B}=&\frac{\Delta_1}{2}(\Id- \opz{B}{1})+\frac{\Delta_2}{2}(\Id- \opz{B}{2})\nonumber\\
& + \frac{\sqrt{\gamma_1 \gamma_2}}{2i}\left (\oppl{B}{1}  \opmi{B}{2} - \oppl{B}{2} \opmi{B}{1}  \right ).
\end{align}

After substituting the SLH parameters and going through the commutator algebra we have the following differential equations

\begin{widetext}
\begin{subequations} \label{eq:DE2scoun}
    \begin{align}
    \partial_t \opmi{A}{1} & = - \left(\frac{\gamma_1}{2}+i \Delta_1 \right) \opmi{A}{1} - i \chi_1 \opmi{A}{1}(\Id-\opz{B}{1}) - \sqrt{\gamma_1}\opz{A}{1} \opin{a}(t), \label{eq:DE2scouna} \\
    \partial_t \opmi{B}{1} & = - \left(\frac{\gamma_1}{2}+i \Delta_1 \right) \opmi{B}{1} - i \chi_1 (\Id-\opz{A}{1})\opmi{B}{1} - \sqrt{\gamma_1 \gamma_2}\opz{B}{1} \opmi{B}{2} - \sqrt{\gamma_1}\opz{B}{1} \opin{b}(t), \label{eq:DE2scounb} \\
    \partial_t \opmi{A}{2} & = - \left(\frac{\gamma_2}{2}+i \Delta_2 \right) \opmi{A}{2} - i \chi_2 \opmi{A}{2}(\Id-\opz{B}{2}) - \sqrt{\gamma_1 \gamma_2}\opz{A}{2} \opmi{A}{1} -\sqrt{\gamma_2}\opz{A}{2} \opin{a}(t), \label{eq:DE2scounc} \\
    \partial_t \opmi{B}{2} & = - \left(\frac{\gamma_2}{2}+i \Delta_2 \right) \opmi{B}{2} - i \chi_2 (\Id-\opz{A}{2})\opmi{B}{2}  -\sqrt{\gamma_2} \opz{B}{2} \opin{b}(t), \label{eq:DE2scound} 
     \end{align}
\end{subequations} 
\end{widetext}
and corresponding input-output relations
\begin{subequations}
\begin{align}    
    \opout{a}(t) & = \opin{a}(t) + \sqrt{\gamma_1} \opmi{A}{1} +\sqrt{\gamma_2} \opmi{A}{2}, \label{eq:DE2scoune} \\
    \opout{b}(t) & = \opin{b}(t) + \sqrt{\gamma_1} \opmi{B}{1} +\sqrt{\gamma_2} \opmi{B}{2}. \label{eq:DE2scounf} 
     \end{align}
\end{subequations} 
Notice that the primary difference between the differential equations in this and the previous case is that, here, the equation for $\opmi{B}{1}$ depends on the outcome of the equation for $\opmi{B}{2}$, whereas in the co-propagating case this relation was reversed. This gives rise to a different order of operations when solving the two-photon transport problem, and has fundamental physical implications.

\subsection{Single-photon S-matrix}

By comparing \cref{eq:DE2scopa} and \cref{eq:DE2scopc} with \cref{eq:DE2scouna} and \cref{eq:DE2scounc} in the single-photon limit, it is obvious that the single-photon solutions are the same in both the co- and counter-propagating cases, as one would expect. We will not repeat the calculations here, just quote the result:

\begin{equation} \label{eq:1p2scounscatt2}
\braket{\omega_a ^{-}|\nu_a^{+}}  = \frac{\conj{\Gamma_2}(\omega_a ) \conj{\Gamma_1}(\omega_a )}{\Gamma_2(\omega_a )\Gamma_1(\omega_a )} \delta(\omega_a  - \nu_a).
\end{equation}

\subsection{Two-photon S-matrix}

The relevant scattering matrix element can be written in the same way as as \cref{eq:2p2scopscatt}, since the two input-output relations \cref{eq:DE2scopf} and \cref{eq:DE2scounf} are the same. To avoid further repetition, we will defer all calculations in this section to \Cref{apx:twosystemcoun2p}. 

However we wish to emphasize one important difference between the co-propagating and counter-propagating analyses. In the co-propagating case we had to solve the equation for $\bra{0} \opmi{B}{1} \ket{\nu_a^{+}}$ and use that result to solve for $\bra{0} \opmi{B}{2} \ket{\nu_a^{+}}$.  In the counter-propagating case it is the other way around, as \cref{eq:DE2scounb} is the one that depends on the outcome of \cref{eq:DE2scound}, which is just the natural manifestation of the counter-propagating cascading of the interaction sites. 

\begin{figure}[h]
\includegraphics[width=1\columnwidth]{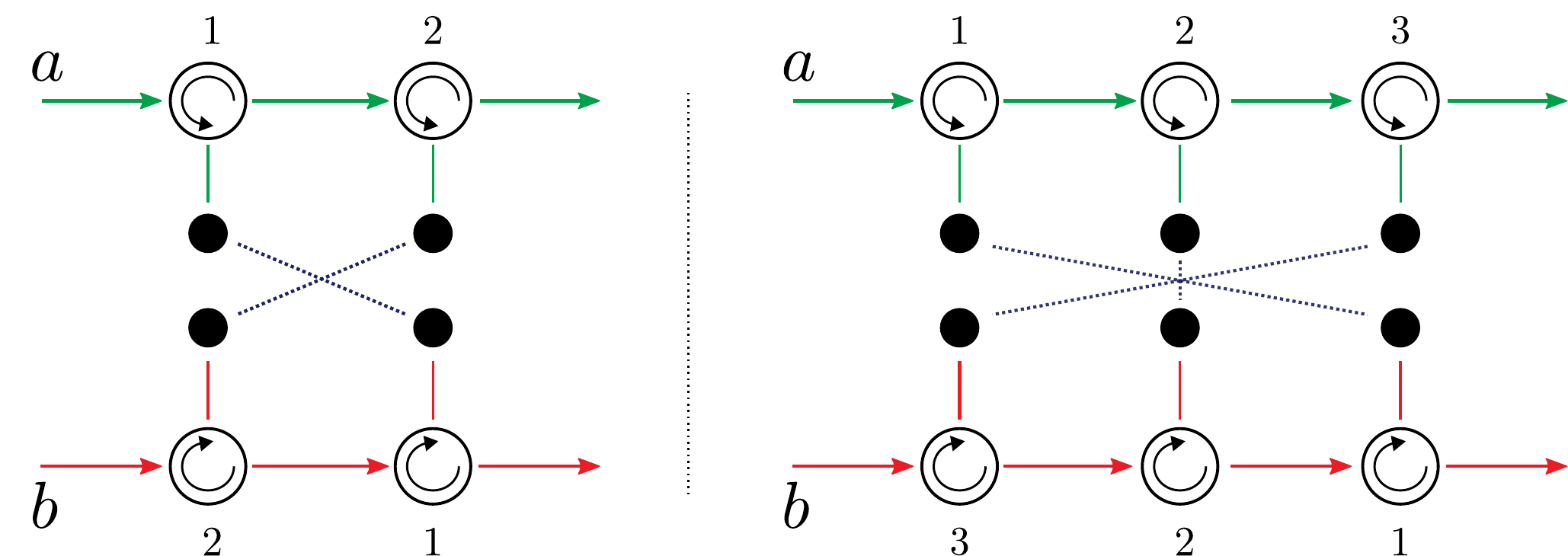}
\caption{Representing a counter-propagating network in a co-propagating manner, we see that local interactions in the former correspond to very nonlocal interaction in the latter. }\label{fig7}
\end{figure}

This seemingly small difference has a major physical implication: while solving the differential equation for $\bra{0} \opmi{B}{1} \ket{\nu_a^{+}}$, the term proportional to $\chi_1 \chi_2$, which would correspond to scattering channel (IV) where the photons interact at both sites, is zero. In fact, the final scattering matrix can be written as 
\begin{widetext}
\begin{align}
\braket{\omega_a ^{-} \omega_b^{-}|\nu_a^+ \nu_b^+}  = & \braket{\omega_a ^{-}|\nu_a^{+}}\braket{\omega_b^{-}|\nu_b^{+}} -  \frac{\delta(\omega_a  + \omega_b - \nu_a - \nu_b)}{\pi} \notag \\ 
& \times \bigg[  i \left( \frac{\conj{\Gamma}_2(\omega_a) \conj{\Gamma}_2(\nu_b)}{\Gamma_2(\omega_a) \Gamma_2(\nu_b)} \right) \left( 1 + \frac{2 i \chi_1}{\Gamma_1(\omega_a) + \Gamma_1(\omega_b)} \right)^{-1} \frac{\chi_1 \gamma_1^2}{\Gamma_1(\nu_b)\Gamma_1(\nu_a) \Gamma_1(\omega_b) \Gamma_1(\omega_a )} \notag \\
& \hphantom{\times \bigg[} + i \left( \frac{\conj{\Gamma}_1(\nu_a) \conj{\Gamma}_1(\omega_b)}{\Gamma_1(\nu_a) \Gamma_1(\omega_b)} \right) \left( 1 + \frac{2 i \chi_2}{\Gamma_2(\omega_a) + \Gamma_2(\omega_b)} \right)^{-1} \frac{\chi_2 \gamma_2^2}{\Gamma_2(\nu_b)\Gamma_2(\nu_a) \Gamma_2(\omega_b) \Gamma_2(\omega_a )} \bigg].
 \label{eq:2p2scounfinal}
\end{align}
\end{widetext}
As we can see from \cref{eq:2p2scounfinal}, the final scattering matrix indeed can be interpreted as the sum of only three channels, corresponding to (I) no interaction, (II) interaction as the first site, or (III) interaction at the second site. This is a generic feature of using counter-propagating wave packets, as we will see in the next session when we write the solution for the $N$-site chain: there are no terms involving interactions at more than one site. An intuitive explanation for this critical difference between the counter- and co-propagating cases can be given by redrawing the atom chains in both cases as if the photons were co-propagating, as in \Cref{fig7}. There we see that, in the counter-propagating case, interactions that happen at more than one site are not {\em causal}, and thus are suppressed from \cref{eq:2p2scounfinal}.  For example, if the two photons interact at site 1, they cannot later (in the timeline of the $a$ photon) interact at site 2, because the photon $b$ has already left that site behind by the time it arrives at 1.  This, of course, gives rise to very different phenomenology in each case, as shown in the companion paper \cite{BrodComb16b}, and as we will discuss in some detail below.

\section{\texorpdfstring{$N$}{N}-site scattering with counter-propagating photons}\label{sec:Ncounter}

Let us now consider the scattering problem for $N$ interaction sites with counter-propagating photons. From the physical intuition developed in the previous section (and our analysis of the three-site interaction, not presented in this manuscript), we have an obvious candidate for the solution: the scattering matrix should be the sum of $N+1$ scattering channels, the first where no interaction happens, and the remaining where the photons interact at the $j$th site, for $1 \leq j \leq N$, and just pick up single-photon phases at the remaining $N-1$ sites. We prove that this is the case, under the simplifying assumption that all interactions site are the same, i.e.\ for all $1 \leq i \leq N$ we have $\gamma_i = \gamma$, $\Delta_i = \Delta$, $\chi_i = \chi$.

\subsection{Cascading and Differential equations}
Here we generalize the counter-propagating analysis from \Cref{sec:2counter}.
The right-going mode (mode $a$) and the left-going mode (mode $b$) are, respectively, represented by the following cascaded systems
\begin{align*}
G_{\rm A}&= \left ( G_{\rm A}^{(N)} \lhd \ldots \lhd G_{\rm A}^{(2)}\lhd G_{\rm A}^{(1)} \right ),\\
G_{\rm B}&= \left ( G_{\rm B}^{(1)} \lhd \ldots \lhd G_{\rm B}^{(N-1)}\lhd G_{\rm B}^{(N)} \right ).
\end{align*}
The total system is then the concatenation of those systems
\begin{align*}
G_{\rm sys}^{\rm counter} 
&= G_{\rm A}\boxplus G_{\rm B}.
\end{align*}
The unwieldy SLH parameters are given in \Cref{apx:nsystemcounSLH}. The resulting differential equations are
\begin{subequations} \label{eq:nsystemDE}
\begin{align}
\partial_t \opmi{A}{k}   =& -\left (\frac{\gamma}{2} +i\Delta \right)\opmi{A}{k}  -i \chi \opmi{A}{k} (\Id-\opz{B}{k})\nonumber\\
& -\gamma \opz{A}{k} \sum_{j=1}^{k-1}\opmi{A}{j} -\sqrt{\gamma}  \opz{A}{k} \opin{a}(t) ,\label{eq:nsystemDEa} \\
\partial_t \opmi{B}{k}   =& -\left (\frac{\gamma}{2} +i\Delta \right)\opmi{B}{k}  -i \chi  (\Id-\opz{A}{k})\opmi{B}{k}\nonumber\\
& -\gamma  \opz{B}{k} \sum_{j=k+1}^{N}\ \opmi{B}{j} -\sqrt{\gamma}  \opz{B}{k} \opin{b}(t), \label{eq:nsystemDEb}
\end{align}
\end{subequations}
and the input-output relations are
\begin{align}
\opout{a}(t) &= \opin{a}(t)+ \sqrt{\gamma}\sum_{j=1}^{N}  \opmi{A}{j}, \\
\opout{b}(t) &= \opin{b}(t)+ \sqrt{\gamma}\sum_{j=1}^{N}  \opmi{B}{j} ,
\end{align} 
as expected.

\subsection{Single-photon S-matrix}

The single-photon S-matrix can now be written as

\begin{align}
\braket{\omega_a ^{-}|\nu_a^{+}} = & \delta(\omega_a  - \nu_a) \notag \\
& + \sqrt{\frac{\gamma}{2\pi}} \sum_{j=1}^{N} \int dt \bra{0} \opmi{A}{j} \ket{\nu_a^{+}} e^{i \omega_a  t}. \label{eq:1pnscopscatt}
\end{align} 

In the \Cref{apx:nsystemcoun1p} we describe how to obtain the solution for this S-matrix from the relevant differential equations. The main new ingredient compared to previous sections is an induction step: we propose a form for the $k$th partial sum 
$\sum_{j=1}^k \bra{0} \opmi{A}{j} \ket{\nu_a^{+}}$, and it can be shown that this form holds for $k=1$ and that if it holds for $k$ then it also holds for $k+1$. Thus, we obtain the solution for \cref{eq:1pnscopscatt} by setting $k=N$, which results in 
\begin{equation} \label{eq:1pnscopscattfinal}
\braket{\omega_a ^{-}|\nu_a^{+}}  = \left(- \frac{\conj{\Gamma}(\omega_a)}{\Gamma(\omega_a)} \right)^N \delta(\omega_a  - \nu_a).
\end{equation}
Again, this just corresponds to each photon picking up the same phase at each of the $N$ sites, as expected.\\

\subsection{Two-photon S-matrix}
Once more, using the single-photon result, we have that 
\begin{align} 
\braket{\omega_a ^{-} \omega_b^{-}|\nu_a^+ \nu_b^+} = & \left(- \frac{\conj{\Gamma}(\omega_a )}{\Gamma(\omega_a )} \right)^N \bigg( \delta(\omega_a  - \nu_a) \delta(\omega_b - \nu_b) \notag \\
& + \sqrt{\frac{\gamma}{2\pi}}  \sum_{j=1}^{N} \int dt e^{i \omega_b t} \bra{\omega_a ^+} \opmi{B}{j} \ket{\nu_a^+ \nu_b^+} \bigg). \label{eq:2p2scounscatt}
\end{align}

Again, we make use of inductive reasoning to obtain the final expression. But now, the indices in the inductive reasoning run ``backwards'' ---we propose a general form for the $m$th partial sum (starting from the end of the chain up until the $m$th site), then it can be shown that this form holds for $m=N$ and that if it holds for $m$ it holds for $m-1$. By setting $m=1$ we obtain our final result. For the explicit expressions for the inductive step we refer the reader to \Cref{apx:nsystemcoun2p}. The final result is:
\begin{widetext}
\begin{align}
\braket{\omega_a ^{-} \omega_b^{-}|\nu_a^+ \nu_b^+}  = & \braket{\omega_a^{-}|\nu_a^{+}} \braket{\omega_b^{-}|\nu_b^{+}} - i \frac{\chi \gamma^2}{\pi} \left(1 + \frac{2 i \chi}{\Gamma(\nu_b) + \Gamma(\nu_a)}\right)^{-1} \frac{\delta(\omega_a + \omega_b - \nu_a - \nu_b)}{\Gamma(\nu_b)\Gamma(\nu_a) \Gamma(\omega_b) \Gamma(\omega_a )} \notag \\
& \hphantom{\braket{\omega_a^{-}|\nu_a^{+}} \braket{\omega_b^{-}|\nu_b^{+}}} \times \bigg[ \sum_{j=1}^{N}
\left( \frac{\conj{\Gamma}(\omega_a)\conj{\Gamma}(\nu_b)}{\Gamma(\omega_a)\Gamma(\nu_b)} \right)^{N-j}
\left( \frac{\conj{\Gamma}(\omega_b)\conj{\Gamma}(\nu_a)}{\Gamma(\omega_b)\Gamma(\nu_a)} \right)^{j-1} \bigg].
\label{eq:2pNscounfinal}
\end{align}
\end{widetext}
This confirms our previous intuition: the first term corresponds to the no-interaction channel, and the sum over $j$ corresponds to the photons interacting at site $j$ and picking up single-photon phases from all other $N-1$ sites other than $j$.

\section{\texorpdfstring{$N \rightarrow \infty$}{N} limit}\label{sec:continuumcounter}
In this Section, we will analyze the $N \rightarrow \infty$ limit of the $N$-site two-photon S-matrix in the counter-propagating arrangement. There are many ways to take this limit, for example we could follow Refs.~\cite{HushCarvHedg13,StevHushCarv14} and take the limit $N\rightarrow\infty$ in the cascaded model, obtain the relevant differential equations, and then calculate the S-matrix. Here we will follow the more direct route of applying the limit directly to the expression of \cref{eq:2pNscounfinal}. 

We begin by noticing that the $\conj{\Gamma}(\omega) / \Gamma(\omega)$ factor which appears at several places in \cref{eq:2pNscounfinal} is simply a phase, so we define the shorthands
\begin{align}
\frac{\conj{\Gamma}(\omega_a)\conj{\Gamma}(\nu_b)}{\Gamma(\omega_a)\Gamma(\nu_b)} = e^{2 i \phi_1} \notag \\
\frac{\conj{\Gamma}(\omega_b)\conj{\Gamma}(\nu_a)}{\Gamma(\omega_b)\Gamma(\nu_a)} = e^{2 i \phi_2}.
\label{eq:phaseshorthands}
\end{align}
Now notice that the sum over $j$ in \cref{eq:2pNscounfinal} is of the form
\begin{equation*}
\sum_{j=1}^{N} x^{N-j} y^{j-1} = \frac{x^N - y^N}{x-y},
\end{equation*}
which in our case leads to
\begin{align}
\sum_{j=1}^{N} (e^{2 i \phi_1} )^{N-j} (e^{2 i \phi_2})^{j-1} =  e^{i (N-1)(\phi_1 + \phi_2)} \frac{\sin[N(\phi_1- \phi_2)]}{\sin[(\phi_1- \phi_2)]}.
\label{eq:phasesum}
\end{align}
The explicit expression for $\phi_1$ is
\begin{equation*}
\phi_1 = \tan^{-1}\left[\frac{2(\omega_a - \Delta)}{\gamma}\right]+\tan^{-1}\left[\frac{2(\nu_b - \Delta)}{\gamma}\right],
\end{equation*}
with a similar equation for $\phi_2$. Let us now make the further assumptions that the input wave packets are spectrally very narrow (compared to $\gamma$), and on resonance with the atoms (i.e.\ $\omega \simeq \Delta$ for all relevant frequencies), such that we can make the approximation $\tan^{-1}(x)\approx x + O(x^3)$, so that:
\begin{align}
\phi_1 \simeq \frac{2}{\gamma}(\omega_a + \nu_b - 2 \Delta), \cr
\phi_2 \simeq \frac{2}{\gamma}(\omega_b + \nu_a - 2 \Delta).
\label{eq:phiapprox}
\end{align}
In the same way, then, we can make the approximation $\sin(\phi_1-\phi_2) \simeq (\phi_1-\phi_2)$ in (\ref{eq:phasesum}), and then, in the limit of large $N$, use the relation $\lim_{N \rightarrow \infty} \sin(N x)/x = \pi \delta(x)$, so
\begin{align*}
\sum_{j=1}^{N} (e^{2 i \phi_1} )^{N-j} (e^{2 i \phi_2})^{j-1} \simeq \pi e^{i (N-1)(\phi_1 + \phi_2)} \delta(\phi_1 - \phi_2).
\end{align*}
From this and (\ref{eq:phiapprox}) we write
\begin{equation}
\delta(\phi_1 - \phi_2) \simeq \frac{\gamma}{2} \delta(\omega_a - \omega_b - \nu_a + \nu_b).
\label{eq:theseconddelta}
\end{equation}
There is an interesting physical interpretation for the appearance of this delta function in the $N \rightarrow \infty$ limit of the S-matrix. In the same way that $\delta(\omega_a + \omega_b - \nu_a - \nu_b)$ can be interpreted as {\em energy} conservation, the $\delta(\omega_a - \omega_b - \nu_a + \nu_b)$ function can be interpreted as {\em momentum} conservation, that arises naturally from the translation-invariance of the medium in the $N \rightarrow \infty$ limit. This also marks a crucial difference between the co- and counter-propagating configurations. In the co-propagating case, energy and momentum conservation lead to the same conditions between the input and output frequencies, and the spectral-entanglement-inducing term seems to be unavoidable. However, in the counter-propagating case, they lead to two distinct conditions, embodied in the two delta functions discussed above. But now we use the following property
\begin{equation}
\delta(\omega_a + \omega_b - \nu_a - \nu_b)\delta(\omega_a - \omega_b - \nu_a + \nu_b) = \frac{1}{2} \delta(\omega_a - \nu_a)\delta(\omega_b - \nu_b), 
\label{eq:deltaproduct}
\end{equation}
to write the final S-matrix as
\begin{widetext}
\begin{align}
\braket{\omega_a ^{-} \omega_b^{-}|\nu_a^+ \nu_b^+}  = & \braket{\omega_a^{-}|\nu_a^{+}} \braket{\omega_b^{-}|\nu_b^{+}} \left[ 1 - i \frac{\chi \gamma^3}{4} \left(1 + \frac{2 i \chi}{\Gamma(\nu_b) + \Gamma(\nu_a)}\right)^{-1} \frac{1}{|\Gamma(\omega_b)\Gamma(\omega_a)|^2}\right].
\label{eq:inftylimitSfinal}
\end{align}
\end{widetext}
This shows that the combination of energy and momentum conservation conditions has the remarkable property of eliminating spectral entanglement altogether.

It may seem that the apparition of this second delta function in  \cref{eq:deltaproduct} contradicts the claim in \cite{XuRephFan13} that, for a finite-size interaction region, the density matrix can contain only one delta function, associated with the conservation of the total energy.  We note that the proof in \cite{XuRephFan13} explicitly assumes a copropagating geometry, and would have to be modified to deal with the counterpropagating case. Even so, there is technically no direct contradiction with that result because, for finite $N$, our second delta function in \cref{eq:deltaproduct}, as we have seen, is only an approximation, albeit one that becomes better and better as the number of sites (and hence the size of the interaction region) increases.  The important point, however, is that even only an approximate delta function can be very useful, as the numerical results in Ref.~\cite{BrodComb16b} demonstrate.

Mathematically, the derivation given above for this second ``delta'' function in \cref{eq:deltaproduct} shows that for a finite number of sites $N$ its actual width is of the order of $1/N$ in the variable $\phi$, or about $\gamma/N$ in frequency units, which could be called the bandwidth of the ``momentum-conservation'' condition for our scheme.

We can make one final simplification to \cref{eq:inftylimitSfinal}, under the assumption that $\omega \simeq \Delta$, to replace $\Gamma(\omega)$ by $\gamma/2$ and obtain
\begin{align}
\braket{\omega_a ^{-} \omega_b^{-}|\nu_a^+ \nu_b^+}  = & \braket{\omega_a^{-}|\nu_a^{+}} \braket{\omega_b^{-}|\nu_b^{+}} \left( \frac{\gamma - 2 i \chi}{\gamma + 2 i \chi} \right) \nonumber\\
= & \braket{\omega_a^{-}|\nu_a^{+}} \braket{\omega_b^{-}|\nu_b^{+}} e^{ -i2\tan^{-1}\left( { 2 \chi}/{\gamma} \right)}.
\end{align}
Finally, by taking the limit that $\chi \rightarrow \infty$, we obtain
\begin{align}
\braket{\omega_a ^{-} \omega_b^{-}|\nu_a^+ \nu_b^+}  = & - \braket{\omega_a^{-}|\nu_a^{+}} \braket{\omega_b^{-}|\nu_b^{+}}.
\end{align}
This shows that, under these approximations, two (spectrally) unentangled photons input in the medium will remain (spectrally) unentangled when they exit it, and acquire an overall (-1) phase beyond whatever phase is induced by the individual single-photon propagators. This is exactly the condition required for a perfect CPHASE gate, which would allow for fully-optical quantum computation. Indeed, in a companion paper to this one \cite{BrodComb16b}, two of us have shown how to obtain a high-fidelity CPHASE gate in this setup, and studied how this fidelity improves as a function of the number of interaction sites (and wave packet bandwidth). Perhaps surprisingly, it is shown there that twelve sites are sufficient to obtain a CPHASE gate with a fidelity of $>99\%$.

\section{Conclusions}\label{sec:conclusion}
In this paper, we considered a fundamental problem in quantum nonlinear optics, namely few photon transport in an atom-mediated $\chi^{(3)}$ medium. Specifically, we computed the S-matrix for single- and two-photon transport through discrete 1D Kerr media in several situations. We considered a single atomic site composed of two coupled atoms; two cascaded atomic interaction sites, with the photons co- or counter-propagating; a cascaded chain of $N$ interaction sites, with counter-propagating photons; and, finally, we studied the double limit of $N \rightarrow \infty$ and narrow input wave packets.

The main difference we identified between co- and counter-propagating cases is the additional term in the co-propagating S-matrix, proportional to $\chi_1 \chi_2$ in \cref{eq:2p2scopfinal}, which corresponds to an interaction between photons at more than one site.
Beyond the two-site case, the presence of this term in \cref{eq:2p2scopfinal} indicates that the scattering solution for the $N$-site co-propagating case would be more complicated than the counter-propagating case, i.e.\ \cref{eq:2pNscounfinal}. In particular, we would expect that the full scattering matrix would contain all terms proportional to $\chi$, $\chi^2$, $\chi^3$ \ldots $\chi^N$, corresponding to interactions happening at 1, 2, 3 \ldots $N$ sites respectively. 

The differences between counter- and co-propagating photons might not manifest for all parameter regimes. In particular, it is possible that, when the $\chi_i$'s and or $\gamma_i$'s are sufficiently small, the expression for the $N$-site co-propagating S-matrix reduces to the one from the counter-propagating case, indicating that the problem can be treated in a perturbative approach. An apparent example of this, in the context of waves propagating in spin chains, is a recent paper by Thompson \etal~\cite{ThomGoklLloy15}, where the authors consider a perturbative analysis of two co-propagating wave packets in a configuration that is similar in spirit to ours. Another example of this limit might be the work of Chudzicki \etal~\cite{ChudChuaShap13}, where the authors explicitly choose to work in a regime in which the phase shift acquired at each site is very small, and the total phase shift of $\pi$ would have to be accumulated over many sites.  However, in their scheme the (undesirable) spectral entanglement would also accumulate from site to site, and the only way to get a high fidelity gate is to make use of a series of projective measurements interspersed with the nonlinear interactions.  This is in contrast with our counterpropagating-pulse scheme, in which we get the entirety of the phase shift from just one site, and the role of the other sites is, essentially, to \emph{passively} remove the spectral entanglement by generating (through interference) the ``momentum-conservation'' delta function i.e. \cref{eq:theseconddelta}.

Our work is not the first to consider the use of counter-propagating wave packets for the implementation of a CPHASE gate, and other examples include \cite{MassFlei04,FriePetrFlei05,HeMacRHan11,HeSche12,BienChoiFirs14,FeizDmocStei16,GorsOtteDeml10,GorsOtteFlei11,HeSharShen14,ShahKuriFlei11}. For example, in \cite{FriePetrFlei05} the authors consider the photons scattering in a counter-propagating manner through an atomic vapour medium, and in \cite{GorsOtteDeml10} the authors couple the fields to a spin chain, and the wave packets counter-propagate as fermionic waves through the spin chain. The main conclusion of these works seems to be that counter-propagating wave packets really undergo more uniform cross-phase modulations than co-propagating ones, leading to a higher-fidelity CPHASE gate between the photons. We believe the work developed here and in our companion paper \cite{BrodComb16b} support these claims, although there are a few important differences from these previous works.

First, we explicitly compute the S-matrices for a $N$-site chain, which leads to a larger flexibility in terms of physical implementations---we can investigate the phenomena as a function of $N$  (and in fact, we do so in \cite{BrodComb16b}) to find out how large $N$ must be for a desired application to work within some quality threshold, but we can also take a $N \rightarrow \infty$ limit and use the result to model a continuum medium such as an atomic vapour. 

Second, our formalism makes it easier to consider a non-translation-invariant medium. Although we only do that explicitly for the two-site case for simplicity, it should not be too hard to generalize our $N$-site analysis for the case where parameters $\gamma_i$, $\Delta_i$ and $\chi_i$ are not all equal---and then one could investigate different longitudinal profiles of these parameters to optimize the operation of a CPHASE gate, or any other task one may wish to perform. For example, consider the work of Hush \etal~\cite{HushCarvHedg13}, which constructs a photonic memory by taking $N\rightarrow \infty$ and choosing $\Delta$ to be a linear ramp in the continuous medium.

Finally, we point out that some of the other proposals with counter-propagating photons, notably \cite{MassFlei04,FriePetrFlei05}, have been criticized (see e.g.\ \cite{HeMacRHan11}) on the grounds that, although the wave packets do not accumulate spectral entanglement during the scattering, they do get entangled in their transversal modes when one considers a fully 3D treatment. However, our approach could also be suitable for a 1D chain of coupled atoms embedded in a waveguide, in which case the 3D analysis of \cite{HeMacRHan11} might not be applicable.

While we have spent some time discussing the application of our results to a CPHASE gate we hope that they will become useful in other quantum nonlinear optics applications. Other applications might require transport for more than one photon per mode. We leave for future work the generalization of our results to multi-photon transport in either a cavity-based Kerr interaction or an equivalent atomic realization.

{\em Acknowledgments:} The authors acknowledge helpful discussions with Agata Bra\'nczyk, Daniel Gottesman, Raissa Mendes, and Zak Webb. The authors also thank Jeffrey Shapiro for feedback and friendly discussions about preliminary versions of this work. J.~C.~ and D.~J.~B.~ acknowledge support by the Perimeter Institute for
Theoretical Physics. Research at Perimeter Institute is supported by the Government of Canada through the Department of Innovation, Science and Economic Development and by the Province of Ontario through the Ministry of Research, Innovation and Science.


\bibliography{../../scattering_ref}


\onecolumngrid
\appendix


\section{Single-site scattering} \label{apx:singlesystem}

The purpose of this and the following appendices is to fill in the gaps of the more technical aspects of the calculations performed in \Crefrange{sec:single}{sec:Ncounter}. In particular, this appendix is concerned with the single- and two-photon scattering in a single-site interaction. Although this is the simplest case, it already contains almost all of the technical ingredients important for the remaining results, and so we describe its steps in more detail. We also point out that some of the key mathematical steps of this proof are similar to those found e.g.\ in Refs. \cite{GeaBNeme14,ViswGeaB15}, although the starting points and formalisms used are quite different.

\subsection{Single-photon transport} \label{apx:singlesystem1p}

As described in \Cref{sec:single}, the elements of the S-matrix can be written as:
\begin{equation*}
\braket{\omega_a ^{-}|\nu_a^{+}} = \frac{1}{\sqrt{2\pi}} \int dt \bra{0} \opout{a} (t) \ket{\nu_a^{+}} e^{i \omega_a  t},
\end{equation*}
which, by the input-output relation, i.e. \cref{eq:DE1sc}, becomes
\begin{equation} \label{eq:apx1p1sscatt}
\braket{\omega_a ^{-}|\nu_a^{+}} = \delta(\omega_a  - \nu_a) + \sqrt{\frac{\gamma}{2\pi}} \int dt \bra{0} A_{-} \ket{\nu_a^{+}} e^{i \omega_a  t},
\end{equation}
where we have used $(1/\sqrt{2\pi}) \int dt \bra{0}\opin{a}(t)\ket{\nu_a^+}e^{i\omega_a  t}=\delta(\omega_a -\nu_a)$.
Now we sandwich \cref{eq:DE1sa} between $\bra{0}$ and $\ket{\nu_a^{+}}$ to obtain
\begin{equation} \label{eq:apx1p1sDEmatrix1}
\partial_t \bra{0} A_{-} \ket{\nu_a^{+}} = - \left(\frac{\gamma}{2} + i \Delta \right) \bra{0} A_{-} \ket{\nu_a^{+}} - i \chi \cancel{\bra{0} A_{-}(\Id - B_{z}) \ket{\nu_a^{+}}} - \sqrt{\gamma} \bra{0} A_{z} \opin{a}(t) \ket{\nu_a^{+}},
\end{equation}
where the two-photon term cancels since we are only interested in single-photon transport for now. Since
\begin{equation*}
\bra{0} A_{z} \opin{a}(t) \ket{\nu_a^{+}}  = \bra{0} \opin{a}(t) \ket{\nu_a^{+}} = \frac{1}{\sqrt{2\pi}} e^{-i \nu_a t},
\end{equation*}
\cref{eq:apx1p1sDEmatrix1} becomes
\begin{equation} \label{eq:apx1p1sDEmatrix2}
\partial_t \bra{0} A_{-} \ket{\nu_a^{+}} = - \left(\frac{\gamma}{2} + i \Delta \right) \bra{0} A_{-} \ket{\nu_a^{+}} - \sqrt{\frac{\gamma}{2\pi}} e^{-i \nu_a t}.
\end{equation}
If we now multiply \cref{eq:apx1p1sDEmatrix2} by $\frac{1}{\sqrt{2\pi}} e^{i \omega_a  t}$ and integrate in $t$ we get
\begin{equation*} 
\frac{1}{\sqrt{2\pi}} \int dt e^{i \omega_a  t} \bra{0} A_{-} \ket{\nu_a^{+}} = - \frac{\sqrt{\gamma}}{\Gamma(\omega_a )} \delta (\omega_a  - \nu_a),
\end{equation*}
where we define the shorthand
\begin{align} 
\Gamma(\omega) := \frac{\gamma}{2} + i (\Delta-\omega).
\end{align}
Finally, plugging this into \cref{eq:apx1p1sscatt} we get 
\begin{equation} \label{eq:apx1p1sscatt2}
\braket{\omega_a ^{-}|\nu_a^{+}} = \left(1-\frac{\gamma}{\Gamma(\omega_a )} \right) \delta(\omega_a  - \nu_a) = - \frac{\conj{\Gamma}(\omega_a )}{\Gamma(\omega_a )} \delta(\omega_a  - \nu_a),
\end{equation}
where we used the identity $\frac{\gamma}{\Gamma(\omega)} - 1 = \frac{\conj{\Gamma}(\omega)}{\Gamma(\omega)}$, which will appear often in the following calculations.

\subsection{Two-photon transport} \label{apx:singlesystem2p}

Now we wish to compute the following quantity
\begin{equation*}
\braket{\omega_a ^{-} \omega_b^{-}|\nu_a^+ \nu_b^+} = \bra{0_a 0_b} \opout{a}(\omega_a )\opout{b}(\omega_b)\ket{\nu_a \nu_b}.
\end{equation*}
We start by inserting a resolution of the identity, $\Id = \ketbra{0_a 0_b}{0_a 0_b}+\int ds_a \ketbra{s_a 0_b}{s_a 0_b}+\int ds_b \ketbra{0_a s_b}{0_a s_b}  +{\rm (two\ photon\ terms)}$, between $\opout{a}(\omega_a)$ and $\opout{b}(\omega_b)$. Given that the result is being acted upon by $\bra{0_a 0_b}$, and that there is not scattering between the $a$ modes and the $b$ modes, it is clear that only the terms proportional to $\ketbra{s_a 0_b}{s_a 0_b}$ contribute to the final expression. Consequently we have (reverting to the abbreviated vacuum notation): 

\begin{equation*}
\braket{\omega_a ^{-} \omega_b^{-}|\nu_a^+ \nu_b^+} = \int ds_a \bra{0} \opout{a}(\omega_a )\ketbra{s_a^+}{s_a^+} \opout{b}(\omega_b)\ket{\nu_a^+ \nu_b^+}.
\end{equation*}
Using the 1-photon result, we get
\begin{equation} \label{eq:apx2p1sscatt}
\braket{\omega_a ^{-} \omega_b^{-}|\nu_a^+ \nu_b^+} = - \frac{\conj{\Gamma}(\omega_a )}{\Gamma(\omega_a )} \left( \delta(\omega_a  - \nu_a) \delta(\omega_b - \nu_b) + \sqrt{\frac{\gamma}{2\pi}} \int dt e^{i \omega_b t} \bra{\omega_a ^+} B_{-} \ket{\nu_a^+ \nu_b^+} \right).
\end{equation}
Now we sandwich \cref{eq:DE1sb} between $\bra{\omega_a ^+}$ and $\ket{\nu_a^+ \nu_b^+}$ to obtain
\begin{equation} \label{eq:apx2p1sDEmatrix3}
\partial_t \bra{\omega_a ^+} B_{-} \ket{\nu_a^{+} \nu_b^+} = - \left(\frac{\gamma}{2} + i \Delta \right) \bra{\omega_a ^+} B_{-} \ket{\nu_a^{+} \nu_b^+} - i \chi \bra{\omega_a ^+} (\Id-A_z) B_{-} \ket{\nu_a^{+} \nu_b^+} - \sqrt{\gamma} \bra{\omega_a ^+} B_{z} \opin{b}(t) \ket{\nu_a^{+} \nu_b^+}.
\end{equation}
But we can write
\begin{equation*}
\bra{\omega_a ^+} B_{z} \opin{b}(t) \ket{\nu_a^{+} \nu_b^+} = \frac{1}{\sqrt{2\pi}} \delta(\omega_a  - \nu_a) e^{-i \nu_b t},
\end{equation*}
since $B_z$ acts trivially on states with excitations only on the $a$ subsystem. We can also write
\begin{align*}
\bra{\omega_a ^+} (\Id-A_{z}) B_{-} \ket{\nu_a^{+} \nu_b^+} & = \int dp_a\bra{\omega_a ^+} \Id - A_z \ketbra{p_a^+}{p_a^+} B_{-} \ket{\nu_a^{+} \nu_b^+} \\
& = 2 \int dp_a\bra{\omega_a ^+}A_{+} \ketbra{0}{0} A_{-} \ketbra{p_a^+}{p_a^+} B_{-} \ket{\nu_a^{+} \nu_b^+},
\end{align*}
where we used that $\Id-A_z = 2 A_{+} A_{-}$ and introduced resolutions of the identity between $A_{+} A_{-}$ and between $A_{-} B_{-}$ (as before, we omit terms which are automatically zero due to photon-number counting considerations) . But, from the single-photon case 
\begin{equation*}
\bra{0} A_{-} \ket{p_a^+} = - \sqrt{\frac{\gamma}{2\pi}} \frac{1}{\Gamma(p_a )} e^{- i p_a t}, 
\end{equation*}
and so
\begin{equation*}
\bra{\omega_a ^+} (\Id-A_{z}) B_{-} \ket{\nu_a^{+} \nu_b^+} = \frac{\gamma}{\pi} \frac{1}{\conj{\Gamma}(\omega_a )} \int dp_a\frac{e^{-i (p_a-\omega_a )t}}{\Gamma(p_a)} \bra{p_a^+} B_{-} \ket{\nu_a^{+} \nu_b^+}.
\end{equation*}
Plugging all back into \cref{eq:apx2p1sDEmatrix3} we obtain
\begin{align} 
\partial_t \bra{\omega_a ^+} B_{-} \ket{\nu_a^{+} \nu_b^+} = & - \left(\frac{\gamma}{2} + i \Delta \right) \bra{\omega_a ^+} B_{-} \ket{\nu_a^{+} \nu_b^+} - i \frac{\chi \gamma}{\pi} \frac{1}{\conj{\Gamma}(\omega_a )} \int dp_a\frac{e^{-i (p_a-\omega_a )t}}{\Gamma(p_a)} \bra{p_a^+} B_{-} \ket{\nu_a^{+} \nu_b^+} \notag \\
& - \sqrt{\frac{\gamma}{2\pi}} e^{- i \nu_b t} \delta(\omega_a  - \nu_a). \label{eq:apx2p1sDEmatrix4}
\end{align}
We will now solve this equation in the Fourier domain. We multiply the whole equation by $\frac{1}{\sqrt{2\pi}} e^{i \omega_b t}$, integrate in $t$, and define $f (\omega_b,\omega_a ):=\frac{1}{\sqrt{2\pi}} \int e^{i \omega_b t} \bra{\omega_a ^+}B_{-}\ket{\nu_a^{+} \nu_b^+}$ (we omit the dependency of $f (\omega_b,\omega_a )$ on $\nu_a$ and $\nu_b$ to simplify the notation). Remark that $f (\omega_b,\omega_a )$ is already what we will need for \cref{eq:apx2p1sscatt}, so it will be unnecessary to invert the Fourier transform. The equation then becomes:

\begin{equation*}
\Gamma(\omega_b) f(\omega_b,\omega_a ) =  - i \frac{\chi \gamma}{\pi} \frac{1}{\conj{\Gamma}(\omega_a )} \int dp_a\frac{f(\omega_b+\omega_a -p_a,p_a)}{\Gamma(p_a)} - \sqrt{\gamma}\delta(\omega_a  - \nu_a)\delta(\omega_b - \nu_b).
\end{equation*}

By replacing $\omega_b \rightarrow \omega_b - \omega_a $ in the previous equation and defining $g(\omega_b, \omega_a ):=f(\omega_b-\omega_a ,\omega_a )$, we get
\begin{equation*}
g(\omega_b,\omega_a ) =  - i \frac{\chi \gamma}{\pi} \frac{1}{\Gamma(\omega_b - \omega_a ) \conj{\Gamma}(\omega_a )} \int dp_a\frac{g(\omega_b,p_a)}{\Gamma(p_a)}- \sqrt{\gamma}\frac{1}{\Gamma(\omega_b - \omega_a )} \delta(\omega_a  - \nu_a)\delta(\omega_b - \omega_a - \nu_b).
\end{equation*}
Note that the integral does not depend on $\omega_a $. So we can define
\begin{equation*}
G(\omega_b) = \int dp_a\frac{1}{\Gamma(p_a)} g(\omega_b,p_a),
\end{equation*}
fixing the dependency of $g(\omega_b,\omega_a )$ on $\omega_a $:
\begin{equation} \label{eq:apx2p1sg}
g(\omega_b,\omega_a ) =  - i \frac{\chi \gamma}{\pi} \frac{1}{\Gamma(\omega_b - \omega_a ) \conj{\Gamma}(\omega_a )} G(\omega_b) - \frac{\sqrt{\gamma}}{\Gamma(\omega_b - \omega_a )} \delta(\omega_a  - \nu_a)\delta(\omega_b - \omega_a  - \nu_b).
\end{equation}
We can now plug this into the definition of $G(\omega_b)$
\begin{equation*}
G(\omega_b) = - i \frac{\chi \gamma}{\pi} G(\omega_b) \int dp_a\frac{1}{\Gamma(\omega_b - p_a) \Gamma(p_a) \conj{\Gamma}(p_a)} - \sqrt{\gamma} \int dp_a\delta(p_a- \nu_a)\frac{\delta(\omega_b - p_a- \nu_b)}{\Gamma(\omega_b-p_a)\Gamma(p_a)}.
\end{equation*}
We can solve the first integral via the residue theorem to obtain
\begin{equation*}
\int dp_a\frac{1}{\Gamma(\omega_b - p_a) |\Gamma(p_a)|^2} = \frac{2 \pi}{\gamma(\gamma + 2 i \Delta - i \omega_b)}.
\end{equation*}
The second integral is simple if one reads it as $\int dp_a\delta(p_a- \nu_a) K(p_a)= K(\nu_a)$ for $K(p_a) = \delta(\omega_b - p_a- \nu_b)/\Gamma(\omega_b-p_a)\Gamma(p_a)$.  Then $G(\omega_b)$ is
\begin{equation*}
G(\omega_b) = - \sqrt{\gamma} \left( 1+\frac{2 i \chi}{\gamma + 2 i \Delta - i \omega_b} \right)^{-1} \frac{\delta(\omega_b - \nu_a - \nu_b)}{\Gamma(\nu_b)\Gamma(\nu_a)}.
\end{equation*}

Plugging this back into \cref{eq:apx2p1sg} we get
\begin{equation*} 
g(\omega_b,\omega_a ) =  i \frac{\chi \gamma \sqrt{\gamma}}{\pi} \left( 1+\frac{2 i \chi}{\gamma + 2 i \Delta - i \omega_b} \right)^{-1} \frac{\delta(\omega_b - \nu_a - \nu_b)}{\Gamma(\nu_b)\Gamma(\nu_a) \Gamma(\omega_b - \omega_a ) \conj{\Gamma}(\omega_a )} -  \frac{\sqrt{\gamma}}{\Gamma(\omega_b - \omega_a )} \delta(\omega_a  - \nu_a)\delta(\omega_b - \omega_a  - \nu_b),
\end{equation*}
and, reverting back to $f(\omega_b,\omega_a )$, we have
\begin{equation*}
f(\omega_b,\omega_a ) =  i \frac{\chi \gamma \sqrt{\gamma}}{\pi} \left( 1+\frac{2 i \chi}{\Gamma(\omega_a )+\Gamma(\omega_b)} \right)^{-1} \frac{\delta(\omega_a  + \omega_b - \nu_a - \nu_b)}{\Gamma(\nu_b)\Gamma(\nu_a) \Gamma(\omega_b) \conj{\Gamma}(\omega_a )} - \frac{\sqrt{\gamma}}{\Gamma(\omega_b)} \delta(\omega_a  - \nu_a)\delta(\omega_b - \nu_b).
\end{equation*}
Finally, plugging this result back into \cref{eq:apx2p1sscatt} we get
\begin{align}
\braket{\omega_a ^{-} \omega_b^{-}|\nu_a^+ \nu_b^+} = & - \frac{\conj{\Gamma}(\omega_a )}{\Gamma(\omega_a )} \bigg[ \delta(\omega_a  - \nu_a) \delta(\omega_b - \nu_b) \left( 1 - \frac{\gamma}{\Gamma(\omega_b)} \right) \notag \\
& \hphantom{- \frac{\conj{\Gamma}(\omega_a )}{\Gamma(\omega_a )} \bigg[} + i \frac{\chi \gamma^2}{\pi} \left( 1+\frac{2 i \chi}{\Gamma(\omega_a )+\Gamma(\omega_b)} \right)^{-1} \frac{\delta(\omega_a  + \omega_b - \nu_a - \nu_b)}{\Gamma(\nu_b)\Gamma(\nu_a) \Gamma(\omega_b) \conj{\Gamma}(\omega_a )} \bigg] \notag \\
 = & \braket{\omega_a ^{-}|\nu_a^{+}}\braket{\omega_b^{-}|\nu_b^{+}} - i \frac{\chi \gamma^2}{\pi} \left( 1+\frac{2 i \chi}{\Gamma(\omega_a )+\Gamma(\omega_b)} \right)^{-1} \frac{\delta(\omega_a  + \omega_b - \nu_a - \nu_b)}{\Gamma(\nu_b)\Gamma(\nu_a) \Gamma(\omega_b) \Gamma(\omega_a )}. \label{eq:apx2p1sfinal}
\end{align}

\section{Two-site scattering with co-propagating photons} \label{apx:twosystemcop}

Both this section and the next follow a similar reasoning to that of \Cref{apx:singlesystem}, only with more cumbersome equations. We will omit much of the repeated discussion, focusing on the particularities of each case.

\subsection{Single-photon transport} \label{apx:twosystemcop1p}

Again, we wish to compute
\begin{equation*}
\braket{\omega_a ^{-}|\nu_a^{+}} = \frac{1}{\sqrt{2\pi}} \int dt \bra{0} \opout{a}(t) \ket{\nu_a^{+}} e^{i \omega_a  t},
\end{equation*}
which, by the input-output relation \cref{eq:DE2scope} becomes
\begin{equation} \label{eq:apx1p2scopscatt}
\braket{\omega_a ^{-}|\nu_a^{+}} = \delta(\omega_a  - \nu_a) + \sqrt{\frac{\gamma_1}{2\pi}} \int dt \bra{0} \opmi{A}{1} \ket{\nu_a^{+}} e^{i \omega_a  t} + \sqrt{\frac{\gamma_2}{2\pi}} \int dt \bra{0} \opmi{A}{2} \ket{\nu_a^{+}} e^{i \omega_a  t}.
\end{equation}

Now we sandwich \cref{eq:DE2scopa} and \cref{eq:DE2scopc} between $\bra{0}$ and $\ket{\nu_a^{+}}$, already preemptively neglecting the two-photon terms:
\begin{subequations} \label{eq:apx1p2sDEmatrix1}
\begin{align}
\partial_t \bra{0} \opmi{A}{1} \ket{\nu_a^{+}} & = - \left(\frac{\gamma_1}{2} + i \Delta_1 \right) \bra{0} \opmi{A}{1} \ket{\nu_a^{+}}- \sqrt{\gamma_1} \bra{0} \opz{A}{1} \opin{a}(t) \ket{\nu_a^{+}},  \label{eq:apx1p2sDEmatrix1a} \\
\partial_t \bra{0} \opmi{A}{2} \ket{\nu_a^{+}} & = - \left(\frac{\gamma_2}{2} + i \Delta_2 \right) \bra{0} \opmi{A}{2} \ket{\nu_a^{+}} - 
\sqrt{\gamma_1 \gamma_2} \bra{0} \opz{A}{2} \opmi{A}{1} \ket{\nu_a^{+}} -\sqrt{\gamma_2} \bra{0} \opz{A}{2} \opin{a}(t) \ket{\nu_a^{+}}. \label{eq:apx1p2sDEmatrix1b}
\end{align}
\end{subequations} 
But we also have that 
\begin{equation*}
\bra{0} \opz{A}{1} \opin{a}(t) \ket{\nu_a^{+}}  = \bra{0} \opin{a}(t) \ket{\nu_a^{+}} = \frac{1}{2\pi} e^{-i \nu_a t} = \bra{0} \opz{A}{2} \opin{a}(t) \ket{\nu_a^{+}},
\end{equation*}
and
\begin{equation*}
\bra{0} \opz{A}{2} \opmi{A}{1} \ket{\nu_a^{+}}  = \bra{0} \opmi{A}{1} \ket{\nu_a^{+}},
\end{equation*}
since $\opz{A}{i}$ acts trivially to the left on the vacuum state. With these facts, \cref{eq:apx1p2sDEmatrix1} become
\begin{subequations} \label{eq:apx1p2sDEmatrix2}
\begin{align}
\partial_t \bra{0} \opmi{A}{1} \ket{\nu_a^{+}} & = - \left(\frac{\gamma_1}{2} + i \Delta_1 \right) \bra{0} \opmi{A}{1} \ket{\nu_a^{+}}- \sqrt{\frac{\gamma_1}{2\pi}} e^{-i \nu_a t}.
 \label{eq:apx1p2sDEmatrix2a} \\
\partial_t \bra{0} \opmi{A}{2} \ket{\nu_a^{+}} & = - \left(\frac{\gamma_2}{2} + i \Delta_2 \right) \bra{0} \opmi{A}{2} \ket{\nu_a^{+}} - 
\sqrt{\gamma_1 \gamma_2} \bra{0} \opmi{A}{1} \ket{\nu_a^{+}} -\sqrt{\frac{\gamma_2}{2\pi}} e^{-i \nu_a t}. \label{eq:apx1p2sDEmatrix2b}
\end{align}
\end{subequations} 
Note that \cref{eq:apx1p2sDEmatrix2a} only depends on matrix elements of $\opmi{A}{1}$, so we can solve it and then feed the result into \cref{eq:apx1p2sDEmatrix2b}. Moving to the Fourier domain, \cref{eq:apx1p2sDEmatrix2a} leads directly to
\begin{equation*} 
\frac{1}{\sqrt{2\pi}} \int dt e^{i \omega_a  t} \bra{0} \opmi{A}{1} \ket{\nu_a^{+}} = - \frac{\sqrt{\gamma_1}}{\Gamma_1(\omega_a )} \delta (\omega_a  - \nu_a),
\end{equation*}
where $\Gamma_i (\omega)$ is defined analogously to $\Gamma (\omega)$ but with parameters pertaining to site $i$, and \cref{eq:apx1p2sDEmatrix2b} in turn leads to 
\begin{align*} 
\frac{1}{\sqrt{2\pi}} \int dt e^{i \omega_a  t} \bra{0} \opmi{A}{2} \ket{\nu_a^{+}} & = - \frac{\sqrt{\gamma_2}}{\Gamma_2(\omega_a )}  \delta (\omega_a  - \nu_a) \left( 1 - \frac{\gamma_1}{\Gamma_1(\omega_a )}\right) \\
& = \frac{\sqrt{\gamma_2}}{\Gamma_2(\omega_a )} \frac{\conj{\Gamma_1(\omega_a )}}{\Gamma_1(\omega_a )} \delta (\omega_a  - \nu_a).
\end{align*}
Plugging all back into \cref{eq:apx1p2scopscatt} we get
\begin{equation} \label{eq:apx1p2scopscatt2}
\braket{\omega_a ^{-}|\nu_a^{+}}  = \frac{\conj{\Gamma_2}(\omega_a ) \conj{\Gamma_1}(\omega_a )}{\Gamma_2(\omega_a )\Gamma_1(\omega_a )} \delta(\omega_a  - \nu_a). 
\end{equation}

\subsection{Two-photon transport} \label{apx:twosystemcop2p}

Again we wish to compute the following quantity
\begin{align} 
\braket{\omega_a ^{-} \omega_b^{-}|\nu_a^+ \nu_b^+} = \frac{\conj{\Gamma_1}(\omega_a )\conj{\Gamma_2}(\omega_a )}{\Gamma_1(\omega_a )\Gamma_2(\omega_a )} \bigg( & \delta(\omega_a  - \nu_a) \delta(\omega_b - \nu_b) + \sqrt{\frac{\gamma_1}{2\pi}} \int dt e^{i \omega_b t} \bra{\omega_a ^+} \opmi{B}{1} \ket{\nu_a^+ \nu_b^+} \notag \\
& +\sqrt{\frac{\gamma_2}{2\pi}} \int dt e^{i \omega_b t} \bra{\omega_a ^+} \opmi{B}{2} \ket{\nu_a^+ \nu_b^+} \bigg). \label{eq:apx2p2scopscatt}
\end{align}
\Cref{eq:DE2scopb} and \cref{eq:DE2scopd} now lead to
\begin{subequations} \label{eq:apx2p2scopDEmatrix3}
    \begin{align}
    \partial_t \bra{\omega_a ^+} \opmi{B}{1} \ket{\nu_a^{+} \nu_b^+}  =&  - \left(\frac{\gamma_1}{2} + i \Delta_1 \right) \bra{\omega_a ^+}  \opmi{B}{1} \ket{\nu_a^{+} \nu_b^+} - i \chi_1 \bra{\omega_a ^+} (\Id- \opz{A}{1})  \opmi{B}{1} \ket{\nu_a^{+} \nu_b^+} \notag \\
&    - \sqrt{\gamma_1} \bra{\omega_a ^+} \opz{B}{1} \opin{b}(t) \ket{\nu_a^{+} \nu_b^+}, \label{eq:apx2p2scopDEmatrix3a} \\
    \partial_t \bra{\omega_a ^+} \opmi{B}{2} \ket{\nu_a^{+} \nu_b^+}  =& - \left(\frac{\gamma_2}{2} + i \Delta_2 \right) \bra{\omega_a ^+}  \opmi{B}{2} \ket{\nu_a^{+} \nu_b^+} - i \chi_2 \bra{\omega_a ^+} (\Id- \opz{A}{2})  \opmi{B}{2} \ket{\nu_a^{+} \nu_b^+} \notag \\ &- \sqrt{\gamma_1 \gamma_2} \bra{\omega_a ^+} \opz{B}{2} \opmi{B}{1} \ket{\nu_a^{+} \nu_b^+}  - \sqrt{\gamma_2} \bra{\omega_a ^+} \opz{B}{2} \opin{b}(t) \ket{\nu_a^{+} \nu_b^+}. \label{eq:apx2p2scopDEmatrix3b}
     \end{align}     
\end{subequations} 
We can make similar simplifications as those leading up to \cref{eq:apx2p1sDEmatrix4}, and use the single-photon results, to obtain

\begin{subequations} \label{eq:apx2p2scopDEmatrix4}
    \begin{align}
  \partial_t \bra{\omega_a ^+} \opmi{B}{1} \ket{\nu_a^{+} \nu_b^+} = & - \left(\frac{\gamma_1}{2} + i \Delta_1 \right) \bra{\omega_a ^+} \opmi{B}{1} \ket{\nu_a^{+} \nu_b^+}  - \sqrt{\frac{\gamma_1}{2\pi}} e^{- i \nu_b t} \delta(\omega_a  - \nu_a) \notag \\
  &  - i \frac{\chi_1 \gamma_1}{\pi} \frac{1}{\conj{\Gamma_1}(\omega_a )} \int dp_a \frac{e^{-i (p_a-\omega_a )t}}{\Gamma_1(p_a)} \bra{p_a^+} \opmi{B}{1} \ket{\nu_a^{+} \nu_b^+}, \label{eq:apx2p2scopDEmatrix4a} \\
    \partial_t \bra{\omega_a ^+} \opmi{B}{2} \ket{\nu_a^{+} \nu_b^+} = & - \left(\frac{\gamma_2}{2} + i \Delta_2 \right) \bra{\omega_a ^+} \opmi{B}{2} \ket{\nu_a^{+} \nu_b^+}  - \sqrt{\gamma_1 \gamma_2} \bra{\omega_a ^+} \opmi{B}{1} \ket{\nu_a^{+} \nu_b^+}  -
    \sqrt{\frac{\gamma_2}{2\pi}} e^{- i \nu_b t} \delta(\omega_a  - \nu_a) \notag \\
    &  - i \frac{\chi_2 \gamma_2}{\pi} \frac{\Gamma_1(\omega_a)}{\conj{\Gamma_1}(\omega_a )\conj{\Gamma_2}(\omega_a )} \int dp_a \frac{\conj{\Gamma}_1(p_a)}{\Gamma_1(p_a) \Gamma_2(p_a)} e^{-i (p_a-\omega_a )t} \bra{p_a^+} \opmi{B}{2} \ket{\nu_a^{+} \nu_b^+}. \label{eq:apx2p2scopDEmatrix4b}
  \end{align}     
\end{subequations} 

Moving to the Fourier domain, and defining $f_i(\omega_b,\omega_a ):=\frac{1}{\sqrt{2\pi}} \int e^{i \omega_b t} \bra{\omega_a ^+}\opmi{B}{i}\ket{\nu_a^{+} \nu_b^+}$ we have
\begin{subequations} \label{eq:apx2p2scopDEmatrix5}
\begin{align}
\Gamma_1(\omega_b) f_1(\omega_b,\omega_a )  = & - i \frac{\chi_1 \gamma_1}{\pi} \frac{1}{\conj{\Gamma_1}(\omega_a )} \int dp_b \frac{1}{\Gamma_1(p_a)} f_1(\omega_b+\omega_a -p_a,p_a) - 
\sqrt{\gamma_1}\delta(\omega_a  - \nu_a)\delta(\omega_b - \nu_b),  \label{eq:apx2p2scopDEmatrix5a}\\
\Gamma_2(\omega_b) f_2(\omega_b,\omega_a )  = & - i \frac{\chi_2 \gamma_2}{\pi} \frac{\Gamma_1(\omega_a)}{\conj{\Gamma_1}(\omega_a )\conj{\Gamma_2}(\omega_a )} \int dp_b \frac{\conj{\Gamma}_1(p_a)}{\Gamma_1(p_a) \Gamma_2(p_a)} f_2(\omega_b+\omega_a -p_a,p_a) \notag \\
& - \sqrt{\gamma_1 \gamma_2} f_1(\omega_a,\omega_b) - \sqrt{\gamma_2}\delta(\omega_a  - \nu_a)\delta(\omega_b - \nu_b). \label{eq:apx2p2scopDEmatrix5b}
\end{align}     
\end{subequations} 

Both equations can be solved by the same method as in the single-system case, of defining $g_i(\omega_b, \omega_a ):=f_i(\omega_b-\omega_a ,\omega_a )$, then defining $G_i(\omega_b)$ to be the integral of $g_i(\omega_b, \omega_a )$ in $\omega_a$, with some multiplying factor. Then we solve a consistency equation to obtain $G_i(\omega_b)$, and with that result find the expressions for $g_i(\omega_b, \omega_a )$ and subsequently $f_i(\omega_b, \omega_a )$. \Cref{eq:apx2p2scopDEmatrix5a} should be solved first, since the result for $ f_1(\omega_b,\omega_a ) $ needs to be fed into \cref{eq:apx2p2scopDEmatrix5b} which then can be solved for $ f_2(\omega_b,\omega_a ) $. We will omit these details, as they are not really enlightening, only display the final conclusion of the calculation, obtained by plugging the results for $f_1(\omega_b,\omega_a )$ and $f_2(\omega_b,\omega_a )$ into \cref{eq:apx2p2scopscatt}:

\begin{align}
\braket{\omega_a ^{-} \omega_b^{-}|\nu_a^+ \nu_b^+}  = & \braket{\omega_a ^{-}|\nu_a^{+}}\braket{\omega_b^{-}|\nu_b^{+}} -  \frac{\delta(\omega_a  + \omega_b - \nu_a - \nu_b)}{\pi} \notag \\
& \times \bigg[  i \left( \frac{\conj{\Gamma}_2(\omega_a) \conj{\Gamma}_2(\omega_b)}{\Gamma_2(\omega_a) \Gamma_2(\omega_b)} \right) \left( 1 + \frac{2 i \chi_1}{\Gamma_1(\omega_a) + \Gamma_1(\omega_b)} \right)^{-1} \frac{\chi_1 \gamma_1^2}{\Gamma_1(\nu_b)\Gamma_1(\nu_a) \Gamma_1(\omega_b) \Gamma_1(\omega_a )} \notag \\
& \hphantom{\times \bigg[}+ i \left( \frac{\conj{\Gamma}_1(\nu_a) \conj{\Gamma}_1(\nu_b)}{\Gamma_1(\nu_a) \Gamma_1(\nu_b)} \right) \left( 1 + \frac{2 i \chi_2}{\Gamma_2(\omega_a) + \Gamma_2(\omega_b)} \right)^{-1} \frac{\chi_2 \gamma_2^2}{\Gamma_2(\nu_b)\Gamma_2(\nu_a) \Gamma_2(\omega_b) \Gamma_2(\omega_a )} \notag \\
& \hphantom{\times \bigg[}+ \left( 1 + \frac{2 i \chi_1}{\Gamma_1(\omega_a) + \Gamma_1(\omega_b)} \right)^{-1} \left( 1 + \frac{2 i \chi_2}{\Gamma_2(\omega_a) + \Gamma_2(\omega_b)} \right)^{-1} \frac{4 \chi_1 \chi_2 \gamma_1^2 \gamma_2^2}{\Gamma_1(\nu_b)\Gamma_1(\nu_a) \Gamma_2(\omega_b) \Gamma_2(\omega_a )} \notag \\
& \hphantom{\times \bigg[}\times \frac{1}{(\Gamma_1(\omega_a) + \Gamma_1(\omega_b))(\Gamma_1(\omega_a) + \Gamma_2(\omega_b))(\Gamma_2(\omega_a) + \Gamma_2(\omega_b))} \bigg].
 \label{eq:apx2p2scopfinal}
\end{align}     

\section{Two-site scattering with counter-propagating photons} \label{apx:twosystemcoun}

\subsection{Single-photon transport} \label{apx:twosystemcoun1p}

By comparing \cref{eq:DE2scopa} and \cref{eq:DE2scopc} with \cref{eq:DE2scouna} and \cref{eq:DE2scounc} in the single-photon limit, it is obvious that the single-photon solutions are the same in both the co- and counter-propagating cases, as one would expect. We will not repeat the calculations here, just quote the result:

\begin{equation} \label{eq:apx1p2scounscatt2}
\braket{\omega_a ^{-}|\nu_a^{+}}  = \frac{\conj{\Gamma_2}(\omega_a ) \conj{\Gamma_1}(\omega_a )}{\Gamma_2(\omega_a )\Gamma_1(\omega_a )} \delta(\omega_a  - \nu_a).
\end{equation}

\subsection{Two-photon transport} \label{apx:twosystemcoun2p}

Let us once more compute the following quantity
\begin{align} 
\braket{\omega_a ^{-} \omega_b^{-}|\nu_a^+ \nu_b^+} = \frac{\conj{\Gamma_1}(\omega_a )\conj{\Gamma_2}(\omega_a )}{\Gamma_1(\omega_a )\Gamma_2(\omega_a )} \bigg(& \delta(\omega_a  - \nu_a) \delta(\omega_b - \nu_b) + \sqrt{\frac{\gamma_1}{2\pi}} \int dt e^{i \omega_b t} \bra{\omega_a ^+} \opmi{B}{1} \ket{\nu_a^+ \nu_b^+} \notag \\
&+\sqrt{\frac{\gamma_2}{2\pi}} \int dt e^{i \omega_b t} \bra{\omega_a ^+} \opmi{B}{2} \ket{\nu_a^+ \nu_b^+} \bigg). \label{eq:apx2p2scounscatt}
\end{align}

From the differential equations for the counter-propagating case, \cref{eq:DE2scounb} and \cref{eq:DE2scound}, we can use the same manipulations as in the previous appendix to obtain the equations in the Fourier domain, for $f_i(\omega_b,\omega_a ):=\frac{1}{\sqrt{2\pi}} \int e^{i \omega_b t} \bra{\omega_a ^+}\opmi{B}{i}\ket{\nu_a^{+} \nu_b^+}$:
 
\begin{subequations} \label{eq:apx2p2scounDEmatrix5}
\begin{align}
\Gamma_1(\omega_b) f_1(\omega_b,\omega_a )  = & - i \frac{\chi_1 \gamma_1}{\pi} \frac{1}{\conj{\Gamma_1}(\omega_a )} \int dp_b \frac{1}{\Gamma_1(p_a)} f_1(\omega_b+\omega_a -p_a,p_a) \notag \\
&  - \sqrt{\gamma_1 \gamma_2} f_2(\omega_a,\omega_b) - \sqrt{\gamma_1}\delta(\omega_a  - \nu_a)\delta(\omega_b - \nu_b),  \label{eq:apx2p2scounDEmatrix5a}\\
\Gamma_2(\omega_b) f_2(\omega_b,\omega_a )  = & - i \frac{\chi_2 \gamma_2}{\pi} \frac{\Gamma_1(\omega_a)}{\conj{\Gamma_1}(\omega_a )\conj{\Gamma_2}(\omega_a )} \int dp_b \frac{\conj{\Gamma}_1(p_a)}{\Gamma_1(p_a) \Gamma_2(p_a)} f_2(\omega_b+\omega_a -p_a,p_a) \notag \\
& - \sqrt{\gamma_2}\delta(\omega_a  - \nu_a)\delta(\omega_b - \nu_b). \label{eq:apx2p2scounDEmatrix5b}
\end{align}     
\end{subequations} 

The main difference between this and the co-propagating case is that, in order to solve \cref{eq:apx2p2scopDEmatrix5}, we solved first \cref{eq:apx2p2scopDEmatrix5a} for $f_1(\omega_b,\omega_a )$, and then used that result to solve \cref{eq:apx2p2scopDEmatrix5b} for $f_2(\omega_b,\omega_a )$. Now, this relation is inverted: the equation for $f_2(\omega_b,\omega_a )$ must be solved first, and its result used to solve the equation for $f_1(\omega_b,\omega_a )$. The steps to solve these equations are essentially the same as in the single-system and in the co-propagating case [cf.\ discussion after \cref{eq:apx2p2scopDEmatrix5}], but there is one seemingly small difference that turns out to be of great physical significance. After solving the equation for $f_2(\omega_b,\omega_a )$ and plugging the result into the equation $f_1(\omega_b,\omega_a )$, at a certain point we must solve the following integral:
\begin{equation} \label{eq:apxintegralzero}
\int dp_a \frac{1}{\Gamma_1(\omega_b - p_a)\Gamma_2(\omega_b - p_a)\conj{\Gamma_1}(p_a)\conj{\Gamma_2}(p_a)}.
\end{equation}
The equivalent step in the co-propagating case led to the term proportional to $\chi_1 \chi_2$ in \cref{eq:apx2p2scopfinal}, which we interpreted as being due to the photons interacting at both the first \textbf{and} at the second site. However, in the counter-propagating case this integral is 0! This can be seen by the fact that the integrand only has poles in the Im$(p_a)>0$ half-plane. Since the integrand decays sufficiently fast for $|p_a| \rightarrow \infty$, we can close the integration contour via an infinite semi-circle on the lower half of the complex plane. Since the functions has no poles in that region, we conclude by the residue theorem that the above integral is 0. The important consequence of this fact is that the final scattering matrix does not contain the scattering channel where interactions happened at both sites, which can be confirmed by the final expression obtained at the end of the above calculation:

\begin{align}
\braket{\omega_a ^{-} \omega_b^{-}|\nu_a^+ \nu_b^+}  = & \braket{\omega_a ^{-}|\nu_a^{+}}\braket{\omega_b^{-}|\nu_b^{+}} -  \frac{\delta(\omega_a  + \omega_b - \nu_a - \nu_b)}{\pi} \notag \\
&\times \bigg[  i \left( \frac{\conj{\Gamma}_2(\omega_a) \conj{\Gamma}_2(\nu_b)}{\Gamma_2(\omega_a) \Gamma_2(\nu_b)} \right) \left( 1 + \frac{2 i \chi_1}{\Gamma_1(\omega_a) + \Gamma_1(\omega_b)} \right)^{-1} \frac{\chi_1 \gamma_1^2}{\Gamma_1(\nu_b)\Gamma_1(\nu_a) \Gamma_1(\omega_b) \Gamma_1(\omega_a )} \notag \\
& \hphantom{\times  \bigg[} + i \left( \frac{\conj{\Gamma}_1(\nu_a) \conj{\Gamma}_1(\omega_b)}{\Gamma_1(\nu_a) \Gamma_1(\omega_b)} \right) \left( 1 + \frac{2 i \chi_2}{\Gamma_2(\omega_a) + \Gamma_2(\omega_b)} \right)^{-1} \frac{\chi_2 \gamma_2^2}{\Gamma_2(\nu_b)\Gamma_2(\nu_a) \Gamma_2(\omega_b) \Gamma_2(\omega_a )}\bigg].
 \label{eq:apx2p2scounfinal}
\end{align}     

\section{\texorpdfstring{$N$}{N}-site scattering with counter-propagating photons} \label{apx:nsystemcoun}

Now we consider the scattering problem for $N$ interaction sites with counter-propagating photons. For simplicity we will work in the assumption that all interactions sites are the same, i.e.\ for all $1 \leq i \leq N$ we have $\gamma_i = \gamma$, $\Delta_i = \Delta$, $\chi_i = \chi$, although we believe that our results should not be hard to generalize otherwise. Since a large fraction of the proof just follows along the same steps as done in the previous appendices, we will just give an outline of the new ingredients that are required in this case, which consist essentially of two proofs by induction. 

\subsection{SLH parameters} \label{apx:nsystemcounSLH}

The total system is then the concatenation of those systems
\begin{align}
G_{\rm sys}^{\rm counter} 
&= G_{\rm A}\boxplus G_{\rm B}= \left ( G_{\rm A}^{(N)}\lhd \ldots  \lhd G_{\rm A}^{(2)}\lhd G_{\rm A}^{(1)} \right )\boxplus  \left ( G_{\rm B}^{(1)} \lhd \ldots \lhd G_{\rm B}^{(N-1)}\lhd G_{\rm B}^{(N)} \right ).
\end{align}
To work out the general rule for cascading $N$ elements we need to know the following

\begin{align}
  G^{(i+1)}_{\rm A} \triangleleft G^{(i)}_{\rm A} &= 
\left( S_{i+1}S_{i}, L_{i+1}+S_{i+1}L_{i}, H_{i}+H_{i+1}+\frac{1}{2i}(L_{i+1}^\dag S_{i+1}L_{i}-L_{i}^\dag S_{i+1}^\dag L_{i+1}) \right),
\end{align}
\begin{align}
  G^{(i)}_{\rm B} \triangleleft G^{(i+1)}_{\rm B} &= 
\left( S_{i}S_{i+1}, L_{i}+S_{i}L_{i+1}, H_{i+1}+H_{i}+\frac{1}{2i}(L_{i}^\dag S_{i}L_{i+1}-L_{i+1}^\dag S_{i}^\dag L_{i}) \right).
\end{align}
Using our model this gives the following 
\begin{align}
 G_{\rm A}&= \left (S_{\rm A}, L_{\rm A} , H_{\rm A} \right),\\
S_{\rm A} &= \Id, \\
L_{\rm A} &=  \sum_{i=1}^N \sqrt{\gamma_i}\opmi{A}{i}, \\
H_{\rm A} &= \sum_{i=1}^N \frac{\Delta_i}{2}(\Id-\opz{A}{i})+ \chi_i (\Id -\opz{A}{i})(\Id -\opz{B}{i}) +
 \frac{1}{2i}\sum_{i=2}^{N}\sum_{j=1}^{i-1} \sqrt{\gamma_{i}\gamma_{j}}\left ( \opmi{A}{i}\phantom{}^\dag \opmi{A}{j}- \opmi{A}{j}\phantom{}^\dag  \opmi{A}{i} \right ),
\end{align}
\begin{align}
G_{\rm B}&= \left (S_{\rm B}, L_{\rm B} , H_{\rm B} \right),\\
S_{\rm B} &= \Id, \\
L_{\rm B} &=  \sum_{i=N}^{1} \sqrt{\gamma_i}\opmi{B}{i},\\
H_{\rm B} &=  \sum_{i=N}^{1} \frac{\Delta_i}{2} (\Id-\opz{B}{i}) +
 \frac{1}{2i} \sum_{i=2}^{N}\sum_{j=1}^{i-1} \sqrt{\gamma_{N+1-i}\gamma_{N+1-j}}\left (  \opmi{B}{N+1-i}\phantom{}^\dag \opmi{B}{N+1-j}- \opmi{B}{N+1-j}\phantom{}^\dag  \opmi{B}{N+1-i} \right ).
\end{align}
The cavity-based Kerr SLH parameters look similar.

\subsection{First inductive step: single-photon transport} \label{apx:nsystemcoun1p}

Again, the expression for the S-matrix elements can be written as
\begin{equation} \label{eq:apx1pnscopscatt}
\braket{\omega_a ^{-}|\nu_a^{+}} = \delta(\omega_a  - \nu_a) + \sqrt{\frac{\gamma}{2\pi}} \sum_{j=1}^{N} \int dt \bra{0} \opmi{A}{j} \ket{\nu_a^{+}} e^{i \omega_a  t}.
\end{equation}
But notice, from \cref{eq:nsystemDEa}, that the differential equation that we need to solve the $k$th term $\bra{0} \opmi{A}{k} \ket{\nu_a^{+}}$ depends on the results for the partial sum of the previous operators from $1$ to $k-1$. So the inductive reasoning is as follows: we will propose a general expression satisfied by the $k$th partial sum, which we obtained by inspection on equivalent results from the $2$- and $3$-site cases (the latter not included in this paper). It can then be proven, which we leave as an exercise for the interested reader, that (i) this expression holds for $k=1$, and (ii) if it holds for $k$ then it also holds for $k+1$. Our final result is then obtained by setting $k = N$, which corresponds to the complete sum over all sites required for \cref{eq:apx1pnscopscatt}. Concretely, the $k$th proposition is:
\begin{equation} \label{eq:apx1pnscopscattpartial}
\delta(\omega_a  - \nu_a) + \sqrt{\frac{\gamma}{2\pi}} \sum_{j=1}^{k} \int dt \bra{0} \opmi{A}{j} \ket{\nu_a^{+}} e^{i \omega_a  t} = (-1)^k \left( \frac{\conj{\Gamma}(\omega_a)}{\Gamma(\omega_a)} \right)^k \delta(\omega_a  - \nu_a).
\end{equation}
After showing that the inductive hypothesis holds, we obtain the final result for $k=N$:
\begin{equation} \label{eq:apx1pnscopscattfinal}
\braket{\omega_a ^{-}|\nu_a^{+}}  =  \left(- \frac{\conj{\Gamma}(\omega_a)}{\Gamma(\omega_a)} \right)^N \delta(\omega_a  - \nu_a).
\end{equation}

\subsection{Second inductive step: two-photon transport} \label{apx:nsystemcoun2p}

Once more, using the single-photon result, we have that 
\begin{equation} 
\braket{\omega_a ^{-} \omega_b^{-}|\nu_a^+ \nu_b^+} =  \left(- \frac{\conj{\Gamma}(\omega_a )}{\Gamma(\omega_a )} \right)^N \bigg( \delta(\omega_a  - \nu_a) \delta(\omega_b - \nu_b) + \sqrt{\frac{\gamma}{2\pi}}  \sum_{j=1}^{N} \int dt e^{i \omega_b t} \bra{\omega_a ^+} \opmi{B}{j} \ket{\nu_a^+ \nu_b^+} \bigg). \label{eq:apx2pnscounscatt}
\end{equation}
From \cref{eq:nsystemDEb}, it is clear that each differential equation for  $\bra{\omega_a^{+}} \opmi{B}{m} \ket{\nu_a^{+} \nu_b^{+}}$ depends on partial sums which now go from the $N$th to the $m+1$th site. We can use the same inductive reasoning as before, but now running ``backwards'': we propose a general form for the $m$th partial sum (starting from the end of the chain), then it can be shown that (i) it holds for $m=N$, and that (ii) if it holds for $m$ then it holds for $m-1$. Finally, we obtain our final result by setting $m=1$. Given $f_i(\omega_b,\omega_a ):=\frac{1}{\sqrt{2\pi}} \int e^{i \omega_b t} \bra{\omega_a ^+}\opmi{B}{i}\ket{\nu_a^{+} \nu_b^+}$, the $m$th proposition is
\begin{align}
\delta(\omega_a - \nu_a)\delta(\omega_b - \nu_b) + \sqrt{\gamma} \sum_{j=m}^{N} f_j(\omega_a, \omega_b) = & \left(- \frac{\conj{\Gamma}(\omega_b )}{\Gamma(\omega_b )} \right)^{N-m+1} \delta(\omega_a - \nu_a)\delta(\omega_b - \nu_b) \notag \\
& + i \frac{\chi \gamma^2}{\pi} (-1)^{N-m} \left(1 + \frac{2 i \chi}{\Gamma(\nu_b) + \Gamma(\nu_a)}\right)^{-1} \frac{\delta(\omega_a + \omega_b - \nu_a - \nu_b)}{\Gamma(\nu_b)\Gamma(\nu_a) \Gamma(\omega_b) \conj{\Gamma(\omega_a )}} \notag \\& \times \frac{\Gamma(\omega_a)^{N-1}\Gamma(\omega_b)^{m-1}}{\conj{\Gamma}(\omega_a)^{N-1}\conj{\Gamma}(\omega_b)^{m-1}}  \sum_{j=m}^{N}
\left( \frac{\conj{\Gamma}(\omega_a)\conj{\Gamma}(\nu_b)}{\Gamma(\omega_a)\Gamma(\nu_b)} \right)^{N-j}
\left( \frac{\conj{\Gamma}(\omega_b)\conj{\Gamma}(\nu_a)}{\Gamma(\omega_b)\Gamma(\nu_a)} \right)^{j-1}.
\label{eq:apx2pnscounpartial}
\end{align}
Again, we leave as an exercise for the reader to check that the requirements for the induction hold. The main aspect of the omitted calculations which is important to point out is that, like in the counter-propagating two-site case, all of the integrals that would lead to terms with higher powers of $\chi$, which would correspond to scattering channels where the photons interact at more than one site, are zero, for the same reasons as \cref{eq:apxintegralzero}. Setting $m=1$ in \cref{eq:apx2pnscounpartial} and plugging into \cref{eq:apx2pnscounscatt} we obtain our final result
\begin{align}
\braket{\omega_a ^{-} \omega_b^{-}|\nu_a^+ \nu_b^+}  = & \braket{\omega_a^{-}|\nu_a^{+}} \braket{\omega_b^{-}|\nu_b^{+}} - i \frac{\chi \gamma^2}{\pi} \left(1 + \frac{2 i \chi}{\Gamma(\nu_b) + \Gamma(\nu_a)}\right)^{-1} \frac{\delta(\omega_a + \omega_b - \nu_a - \nu_b)}{\Gamma(\nu_b)\Gamma(\nu_a) \Gamma(\omega_b) \Gamma(\omega_a )} \notag \\
& \hphantom{\braket{\omega_a^{-}|\nu_a^{+}} \braket{\omega_b^{-}|\nu_b^{+}}} \times \sum_{j=1}^{N} \left( \frac{\conj{\Gamma}(\omega_a)\conj{\Gamma}(\nu_b)}{\Gamma(\omega_a)\Gamma(\nu_b)} \right)^{N-j}
\left( \frac{\conj{\Gamma}(\omega_b)\conj{\Gamma}(\nu_a)}{\Gamma(\omega_b)\Gamma(\nu_a)} \right)^{j-1}.
\label{eq:apx2pnscounfinal}
\end{align}

\end{document}